\documentclass[fleqn,usenatbib]{mnras}

\usepackage[T1]{fontenc}

\DeclareRobustCommand{\VAN}[3]{#2}
\let\VANthebibliography\thebibliography
\def\thebibliography{\DeclareRobustCommand{\VAN}[3]{##3}\VANthebibliography}

\usepackage{graphicx}	
\usepackage{amsmath}	
\usepackage{amssymb}	
\usepackage{siunitx}

\usepackage{filecontents}
\usepackage{subfigure}
\usepackage{tikz}
\usepackage{multirow}
\usepackage{xcolor}
\usepackage{multicol}
\usepackage{hyperref}

\newcommand{\Msun}{\mbox{M$_{\odot}$}}
\newcommand{\MJup}{\mbox{M$_{\mathrm{Jup}}$}}

\newcommand{\Teff}{\mbox{$T_\mathrm{eff}$}}
\newcommand{\kms}{\mbox{km\,s$^{-1}$}}
\newcommand{\EW}{\mbox{$W_{\textnormal{$\lambda$}}$}}

\title[An emerging DAe white dwarf class]{An emerging and enigmatic spectral class of isolated DAe white dwarfs}

\author[A. K. Elms]{Abbigail K. Elms,$^{1}$\thanks{Contact e-mail: \href{mailto:Abbigail.Elms@warwick.ac.uk}{Abbigail.Elms@warwick.ac.uk}}
{Pier-Emmanuel Tremblay,$^{1}$}
{Boris T. G{\"a}nsicke,$^{1}$}
{Andrew Swan,$^{1}$}
{Carl Melis,$^{2}$}
\newauthor
{Antoine B\'edard,$^{1}$}
{Christopher J. Manser,$^{3}$}
{James Munday,$^{1,4}$}
{J. J. Hermes,$^{5}$}
{Erik Dennihy,$^{6}$}
\newauthor
{Atsuko Nitta,$^{7}$ and}
{Ben Zuckerman$^{8}$}
\\
$^{1}$Department of Physics, University of Warwick, Coventry, CV4 7AL, UK\\
$^{2}$Center for Astrophysics and Space Sciences, University of California, San Diego, CA 92093-0424, USA\\
$^{3}$Astrophysics Group, Department of Physics, Imperial College London, Prince Consort Rd, London, SW7 2AZ, UK\\
$^{4}$Isaac Newton Group of Telescopes, Apartado de Correos 368, E-38700 Santa Cruz de La Palma, Spain\\
$^{5}$Department of Astronomy, Boston University, 725 Commonwealth Ave., Boston, MA 02215, USA
\\
$^{6}$Rubin Observatory Project Office, 950 N. Cherry Ave., Tucson, AZ 85719, USA\\
$^{7}$Gemini Observatory, 670 N A'ohoku Pl., Hilo, Hawaii, HI 96720, USA\\
$^{8}$Department of Physics and Astronomy, University of California, Los Angeles, CA 90095-1562, USA\\
}

\date{Accepted 2023 July 17. Received 2023 July 17; in original form 2023 June 16}

\pubyear{2023}

\usepackage{newtxtext,newtxmath}

\begin{document}
\label{firstpage}
\pagerange{\pageref{firstpage}--\pageref{lastpage}}
\maketitle

\begin{abstract}
Two recently discovered white dwarfs, WD\,J041246.84$+$754942.26 and WD\,J165335.21$-$100116.33, exhibit H$\alpha$ and H$\beta$ Balmer line emission similar to stars in the emerging DAHe class, yet intriguingly have not been found to have detectable magnetic fields. These white dwarfs are assigned the spectral type DAe. We present detailed follow-up of the two known DAe stars using new time-domain spectroscopic observations and analysis of the latest photometric time-series data from \textit{TESS} and ZTF. We measure the upper magnetic field strength limit of both stars as $B < 0.05$\,MG. The DAe white dwarfs exhibit photometric and spectroscopic variability, where in the case of WD\,J041246.84$+$754942.26 the strength of the H$\alpha$ and H$\beta$ emission cores varies in anti-phase with its photometric variability over the spin period, which is the same phase relationship seen in DAHe stars. The DAe white dwarfs closely cluster in one region of the \textit{Gaia} Hertzsprung-Russell diagram together with the DAHe stars. We discuss current theories on non-magnetic and magnetic mechanisms which could explain the characteristics observed in DAe white dwarfs, but additional data are required to unambiguously determine the origin of these stars.

\end{abstract}




\section{Introduction}

Almost four decades ago, the isolated white dwarf GD\,356 (WD\,1639$+$537) was discovered and found to be magnetic with a hydrogen-dominated atmosphere and Zeeman-split H$\alpha$ and H$\beta$ Balmer line emission triplets \citep{Greenstein1985}. It was assigned the spectral class DAHe, a degenerate (D) star with Balmer lines (A), Zeeman-splitting (H) and emission (e). GD\,356 was later found to be phometrically variable over a period of 115\,minutes \citep{Brinkworth2004,Wickramasinghe2010}. GD\,356 remained the only member of its class until a few years ago, when \citet{Gansicke2020} and \citet{Reding2020} discovered two new DAHe stars. These new stars evidenced spectroscopic variability in the H$\alpha$ and H$\beta$ emission features in addition to photometric variability. \citet{Walters2021} conducted recent work on GD\,356 with new time-resolved data and also confirmed spectroscopic and photometric variability in this white dwarf. Survey data and targeted spectroscopic follow-up of DAHe candidates, selected due to their photometric variability from hundreds of thousands of white dwarf candidates first identified from the spacecraft \textit{Gaia} \citep{Gaia2022}, have led to the discovery of 26 DAHe stars to date \citep{Manser2023, Reding2023}.

Two interesting white dwarfs, WD\,J041246.84$+$754942.26 \citep[hereafter WD\,J0412$+$7549;][]{Tremblay2020} and WD\,J165335.21$-$100116.33 \citep[hereafter WD\,J1653$-$1001;][]{OBrien2023}, have emerged over the past few years and been classified as DAe~-~they have hydrogen-dominated atmospheres and exhibit weaker H$\alpha$ and H$\beta$ line emission than DAHe stars, but intriguingly lack an observable magnetic field i.e. do not show Zeeman-split emission line triplets. These two stars, in addition to two of the 26 DAHe white dwarfs, lie within 40\,pc of the sun. There are 1066 white dwarfs in the 40\,pc volume-limited sample of white dwarfs \citep{McCleery2020, OBrien2023} of which 655 are classified as DA, so the identification of four DA white dwarfs with Balmer emission lines to date within this volume attests to a fraction of 0.61 per cent. The two DAe and 26 DAHe stars closely cluster in one region on the \textit{Gaia} Hertzsprung–Russell diagram \citep[HRD;][]{Gansicke2020, Walters2021, Manser2023} and have a remarkable homogeneity in atmospheric parameters, with effective temperatures $7400\,\mathrm{K} \lesssim \Teff \lesssim 8500$\,K and white dwarf masses $0.5\,\Msun \lesssim \mathrm{M}_{\mathrm{WD}} \lesssim 0.8$\,\Msun. The DAe and DAHe stars with reliable time-resolved data have measured variability periods of $\simeq 0.08 - 36$\,h, which is plausibly linked to the rotation period.

DAe and DAHe white dwarfs are encapsulated into one DA(H)e class in this work based on our initial hypothesis that these objects have a similar origin but differ in terms of magnetic field strength and possibly other properties. The magnetic field strengths observed at the surface of DAHe stars range from $\simeq 5 - 147$\,MG \citep{Greenstein1985, Gansicke2020, Reding2020, Walters2021, Manser2023, Reding2023}, resulting in a range of physical effects acting upon the white dwarfs including altered or suppressed surface convection and possibly altered hydrostatic structure of atmospheric layers \citep{Landstreet1987,Tremblay2015,Ferrario2015,Fuller2023}. However, $86 \pm ^{7}_{10}$ per cent of magnetic white dwarfs within the same region of the \textit{Gaia} HRD do not show Balmer line emission \citep{Manser2023}, suggesting that magnetic field strength is not the only physical parameter defining the DA(H)e class. Recent analyses of DA(H)e stars show no evidence of binarity or ongoing accretion of planetary debris \citep[e.g. see][for the DAe WD\,J0412$+$7549]{Tremblay2020}. Note that there are instances of DAe+dM systems in the literature where the Balmer emission is related to binarity \citep[e.g.][]{Silvestri2006, Silvestri2007, Gianninas2011}, but these systems are not discussed in this work. 

Previous studies have explored intrinsic (e.g. stellar structure and internal dynamics) and extrinsic (e.g. planetary mass companions and binary interactions) mechanisms to explain the observations of DA(H)e white dwarfs \citep{Gansicke2020, Reding2020, Schreiber2021b, Walters2021, Ginzburg2022, Reding2023}. An active chromosphere which is hosted by a (magnetic) dark surface spot/region is one explanation for Balmer emission lines. The close clustering of DA(H)e objects on the \textit{Gaia} HRD could be explained by a convective dynamo driven by white dwarf core crystallization occurring at a specific time in the cooling sequence \citep{Gansicke2020, Ginzburg2022}, although this scenario has recently been questioned \citep{Fuentes2023}. While no DA(H)e star to date has been found in a binary system, it has been suggested that Balmer line emission could be caused by a planetary mass companion, possibly from magnetic induction in a close-in orbit \citep{Goldreich1969, Li1998, Wickramasinghe2010}. However, no study has been able to unambiguously conclude on a suitable scenario therefore the mechanism(s) causing emission in DA(H)e white dwarfs remains a mystery.

In this work, we present new time-domain spectroscopic observations of the two DAe stars, WD\,J0412$+$7549 and WD\,J1653$-$1001, and analyse their time-series observations from the Transiting Exoplanet Survey Satellite \citep[\textit{TESS};][]{TESS2014} and Zwicky Transient Facility \citep[ZTF;][]{Bellm2019, Masci2019}. In Section~\ref{sec:observations}, we present photometric and spectroscopic time-domain observations of WD\,J0412$+$7549 and WD\,J1653$-$1001. We analyse the data of both white dwarfs in Section~\ref{sec:Analysis} to obtain atmospheric parameters, investigate photometric and spectroscopic line variability and determine limits on radial velocity shifts. In Section~\ref{sec:Discussion}, we discuss our results and conclude in Section~\ref{sec:Conclusions}.

\section{Observations and data}
\label{sec:observations}

\subsection{Broad-band photometry}
\label{sec:Photometry}

\begin{table}
	\centering
	\caption{Optical and IR photometry, including magnitudes from different filters, for WD\,J0412$+$7549 and WD\,J1653$-$1001.}
	\label{tab:photometry}
	\begin{tabular}{llcc}
		\hline
		\hline
		Survey & Filter & WD\,J0412$+$7549 & WD\,J1653$-$1001\\
		& & [mag] & [mag]\\
		\hline
		\hline
		\multirow{3}{2cm}{\textit{Gaia}} & $G$ & 15.815 $\pm$ 0.003 & 15.708 $\pm$ 0.003\\
		& $G_{\rm BP}$ & 15.922 $\pm$ 0.004 & 15.851 $\pm$ 0.005\\
		& $G_{\rm RP}$ & 15.626 $\pm$ 0.005 & 15.385 $\pm$ 0.007\\
		\hline
		\multirow{5}{2cm}{Pan-STARRS} & $g$ & 15.916 $\pm$ 0.008 & 15.872 $\pm$ 0.005 \\
		& $r$ & 15.900 $\pm$ 0.004 & 15.736 $\pm$ 0.004\\
		& $i$ & 15.948 $\pm$ 0.003 & 15.766 $\pm$ 0.001 \\
		& $z$ & 16.047 $\pm$ 0.004 & 15.817 $\pm$ 0.002\\
		& $y$ & 16.124 $\pm$ 0.005 & 15.886 $\pm$ 0.008\\
		\hline
		\multirow{3}{2cm}{2MASS} & $J$ & 15.544 $\pm$ 0.064 & 15.122 $\pm$ 0.057\\
		& $H$ & 15.412 $\pm$ 0.133 & 15.064 $\pm$ 0.086\\
		& $K$ & 15.519 $\pm$ 0.235 & 15.076 $\pm$ 0.138\\
		\hline
	\end{tabular}
\end{table}

\begin{table*}
    \centering
	\caption{Observed and derived parameters of WD\,J0412$+$7549 and WD\,J1653$-$1001. Atmospheric parameters were calculated by performing photometric (Phot) and weighted 3D spectroscopic (3D spec; see text for details) fits. Values are given in the J2016.0 epoch.}
	\label{tab:stellar parameters}
	\begin{tabular}{llccl}
		\hline
		\hline
		Parameter & & WD\,J0412$+$7549 & WD\,J1653$-$1001 & \\
		\hline
		\hline
        Designation & & \textit{Gaia} DR3 551153263105246208 & \textit{Gaia} DR3 4334641562477923712 & \\
		Spectral type & & DAe & DAe & \\
		RA & & 04:12:46.23 & 16:53:35.21 & \\
		Dec & & $+$75:49:42.68 & $-$10:01:16.33 & \\
		Parallax & $\varpi$ [mas] & 28.53 $\pm$ 0.03 & 30.65 $\pm$ 0.04 & \\
		Distance & \textit{d} [pc] & 35.05 $\pm$ 0.04 & 32.63 $\pm$ 0.04 & \\ 
		Proper motion & $\mu_\alpha$ [mas\,yr$^{-1}$] & $-$140.90 $\pm$ 0.03 & 159.38 $\pm$ 0.05 \\	
		 & $\mu_\delta$ [mas\,yr$^{-1}$] & 26.11 $\pm$ 0.03 & $-$211.01 $\pm$ 0.03  & \\
		Absolute magnitude & $M_{\rm G}$ [mag] & 13.092 $\pm$ 0.003 & 13.140 $\pm$ 0.003 & \\
		\hline
		Effective temperature & \Teff\ [K] & 8546 $\pm$ 87 & 7388 $\pm$ 71 & (Phot)\\
		 & & 8578 $\pm$ 106 & 7613 $\pm$ 95 & (3D spec)\\
		Surface gravity & $\log g$ [cm\,s$^{-2}$] &  8.260 $\pm$ 0.030 & 7.930 $\pm$ 0.030 & (Phot)\\
		 & & 8.316 $\pm$ 0.025 & 7.893 $\pm$ 0.030 & (3D spec) \\
		Mass & $\mathrm{M}_{\mathrm{WD}}$ [\Msun] & 0.76 $\pm$ 0.02 & 0.55 $\pm$ 0.02 & (Phot)\\
		 & & 0.80 $\pm$ 0.02 & 0.53 $\pm$ 0.02 & (3D spec) \\
		Radius & $R$ [$\times$10$^{-5}$ R$_\odot$] & 1072 $\pm$ 19 & 1331 $\pm$ 23 & (Phot)\\
		 & & 1027 $\pm$ 19 & 1366 $\pm$ 26 & (3D spec) \\
		Cooling age & $\tau$ [Gyr] & 1.290 $\pm$ 0.057 & 1.154 $\pm$ 0.048 & (Phot)\\
		 & & 1.419 $\pm$ 0.086 & 1.019 $\pm$ 0.050 & (3D spec) \\
		Magnetic field strength & \textit{B} [MG] & < 0.05 & < 0.05 & \\
		Spin period & \textit{P} [h] & 2.2891144 $\pm$ 0.0000016 & 80.534 $\pm$ 0.087 & \\
		\hline
	\end{tabular}
\end{table*}

WD\,J0412$+$7549 and WD\,J1653$-$1001 have photometry in the optical from \textit{Gaia} DR3 and the Panoramic Survey Telescope and Rapid Response System \citep[Pan-STARRS;][]{Chambers2016, Flewelling2020} DR2, in addition to near-IR photometry from the Two Micron All Sky Survey \citep[2MASS;][]{Skrutskie2006}. Table~\ref{tab:photometry} displays the available photometric data. We do not include photometry from the Wide-field Infrared Survey Explorer \citep[\textit{WISE};][]{WISE2010} CatWISE2020 catalogue \citep{Marocco2021} in photometric fits for either object performed in this work as the close proximity of background sources results in the contamination of \textit{WISE} measurements. WD\,J0412$+$7549 is $\approx 15$\, arcsec away from the edge-on dusty galaxy LEDA\,2769388 \citep{Paturel2003} which has an estimated redshift of $z = 0.07$ \citep{Dalya2018}. WD\,J1653$-$1001 is $\approx 1$\,arcsec away from a background main-sequence star (Gaia DR3 4334641562479650816). 

Stellar parameters -- including spectral type, astrometry and atmospheric parameters -- for WD\,J0412$+$7549 and WD\,J1653$-$1001 are shown in Table~\ref{tab:stellar parameters}. Details on the photometric and spectroscopic fits performed to calculate the atmospheric parameters are given in Section~\ref{sec:parameters}.

\subsection{Time-domain spectroscopy}
\label{sec:Spectroscopy}

\begin{table*}
    \centering
	\caption{Time-domain spectroscopic observations for WD\,J0412$+$7549 and WD\,J1653$-$1001 obtained from ground-based telescopes, detailing the exposure time ($t_{\rm exp}$), number of exposures ($n_{\rm exp}$) for each observing run and the duration of the observing run. Numbers separated by a colon represent exposures taken in the blue:red arms.}
	\label{tab:observations}
	\begin{tabular}{cccccc}
		\hline
		\hline
		Object & Date & Telescope/Instrument & $t_{\rm exp}$ & $n_{\rm exp}$ & Duration \\
		&  &  & (s) &  & (h) \\
		\hline
		\hline
		\multirow{8}{2cm}{\textbf{WD\,J0412$+$7549}} & 2018-Oct-14 & WHT/ISIS & 600 & 3 & 0.56 \\
		& 2018-Oct-15 & WHT/ISIS & 600 & 1 & 0.17 \\
		& 2018-Oct-16 & WHT/ISIS & 600 & 1 & 0.17 \\
		& 2019-Dec-09 & Keck/HIRES & 1800 & 2 & 4.29 \\
		& 2020-Aug-17 & INT/IDS & 1200 & 2 & 0.67 \\
		& 2020-Aug-18 & INT/IDS & 900 & 12 & 3.14 \\
		& 2020-Oct-04 & Gemini/GMOS-N & 300 & 8 & 0.73 \\
		& 2020-Oct-05 & Gemini/GMOS-N & 300 & 8 & 0.73 \\
		& 2021-Jan-12 & Gemini/GMOS-N & 300 & 20 & 2.08 \\
		\hline
        \multirow{2}{2cm}{\textbf{WD\,J1653$-$1001}} & 2018-May-22 & Shane/KAST & 3000:1000 & 1:3 & 0.83 \\
        & 2023-May-15 & Shane/KAST & 2000:1000 & 2:4 & 1.11 \\
        \hline
	\end{tabular}
\end{table*}

Spectroscopic observations of WD\,J0412$+$7549 were made using four different ground-based telescopes spanning 27\,months. The long time-frame between observations allows for a dedicated search for variability in the Balmer emission lines. Observational details are listed in Table~\ref{tab:observations} for WD\,J0412$+$7549 and WD\,J1653$-$1001, including the exposure times ($t_{\rm exp}$), number of exposures ($n_{\rm exp}$) and the total duration of each observing run. Sections~\ref{sec:WHT}~--~\ref{sec:Gemini} discuss the observations of WD\,J0412$+$7549 made with each telescope in detail. Section~\ref{sec:KAST} presents the spectroscopic observations of WD\,J1653$-$1001.

\subsubsection{WHT/ISIS}
\label{sec:WHT}
Intermediate-resolution spectroscopy of WD\,J0412$+$7549 was obtained with the double-arm Intermediate-dispersion Spectrograph and Imaging System (ISIS) on the Cassegrain focus of the 4.2-m William Herschel Telescope (WHT) at the Observatorio del Roque de los Muchachos on La Palma, Spain. We used the default CCD detectors EEV12 $2048 \times 4096$\,pixel$^2$ in the blue (R600B grating, resolving power $R \approx 2000$) arm and RED+ $2048 \times 4096$\,pixel$^2$ in the red (R600R grating, $R \approx 2700$) arm. The approximate wavelength ranges covered by the blue and red arms in our observations are $3100$~--~$5400$\,\AA\ and $5700$~--~$9000$\,\AA, respectively, thus all Balmer line (H$\alpha$ to H$\zeta$) regions were observed. We used a slit width of 1.2\,arcsec and dispersions of 0.49\,\AA/pixel in the blue arm and 0.45\,\AA/pixel in the red arm. We imposed a binning of $2 \times 2$, resulting in an average resolution of $\approx 2$\,\AA.

Observations were taken on 2018 October 14-16 with 600\,s exposures. Three exposures were taken on 2018 October 14 while one exposure was taken on each of the subsequent nights.

\subsubsection{Keck/HIRES}
\label{sec:Keck}
High-resolution optical spectra of WD\,J0412$+$7549 were obtained on 2019 December 9 from the High Resolution Echelle Spectrometer \citep[HIRES;][]{Vogt1994} instrument on the Keck-1 10-m telescope at the W. M. Keck Observatory, Hawaii. Observations were taken using the mosaic of three MIT-Lincoln Lab (MIT/LL) $2000 \times 4000$\,pixel$^2$ CCDs with the red collimator (HIRESr) and C5 decker ($R \approx 40\,000$), with slit width 1.148\,arcsec and $1 \times 2$ binning. The wavelength coverage of our HIRES observations is approximately $4800$~--~$6700$\,\AA, with small gaps between echelle orders. The H$\alpha$ and H$\beta$ line regions were covered with this setup with an average resolution of $\approx 0.15$\,\AA.

Two 1800\,s exposures were taken of WD\,J0412$+$7549 with 3.79\,h separation (time between exposure start times) as a check for emission line variability. The spectra were reduced and extracted using the HIRES software package MAKEE\footnote{https://sites.astro.caltech.edu/~tb/makee/}. Keck observed the standard star Feige 110 on the same night as WD\,J0412$+$7549 so we used this spectrum to correct for the instrumental response function (IRF) of the telescope. Although Feige 110 has H$\alpha$ and H$\beta$ lines, the rest of the spectrum is featureless and an accurate representation of the IRF. Therefore, we used $1-3$ orders before/after the Balmer line regions in Feige 110 to correct the spectra of WD\,J0412$+$7549.

\subsubsection{INT/IDS}
\label{sec:INT}
We collected intermediate-resolution spectroscopic observations of WD\,J0412$+$7549 on two consecutive nights using the Intermediate Dispersion Spectrograph (IDS) on the Cassegrain focus of the 2.5-m Isaac Newton Telescope (INT), located at the Observatorio del Roque de los Muchachos on La Palma, Spain. Our setup utilized the EEV10 $4096 \times 2048$\,pixel$^2$ CCD detector in the blue arm with a slit width of $1.2$\,arcsec. We used the R632V grating centred at 5720\,\AA, resulting in a dispersion of 0.90\,\AA/pixel over the approximate wavelength range $4200$~--~$7000$\,\AA\ with a spectral resolution of $R=2400$. We employed $1 \times 1$ binning and have an average resolution of $\approx 2$\,\AA.

Two 1200\,s exposures were taken on 2020 August 17 and twelve 900\,s exposures were taken on 2020 August 18. Spectroscopic coverage was achieved for Balmer lines from H$\alpha$ to H$\gamma$.

\subsubsection{Gemini/GMOS-N}
\label{sec:Gemini}
We used the Gemini Multi-Object Spectrograph \citep[GMOS;][]{Hook2004} instrument on the 8-m Gemini-North Telescope (Gemini-N/GMOS-N) to search for variability in the Balmer line emission of WD\,J0412$+$7549 over multiple epochs as part of programme GN-2020B-Q-304. Long-slit spectroscopic observations were performed using the three GMOS-N $2048 \times 4176$\,pixel$^2$ Hamamatsu CCD \citep{Scharwachter2018} chips. The B600+G5307 grating ($R \approx 1700$) was used with a central wavelength of 5300\,\AA\ and $2 \times 2$ binning. The approximate wavelength range covered by our observations is $3900$~--~$6700$\,\AA. Our setup resulted in a dispersion of 0.45\,\AA/pixel and an average resolution of $\approx 3$\,\AA.

Eight consecutive 300\,s exposures were taken on 2020 October 4 and 5, with an additional observation taken on 2021 January 12 that consisted of 20 consecutive 300\,s exposures. The spectra were flux calibrated and cover Balmer lines from H$\alpha$ to H$\delta$. The additional H$\epsilon$ Balmer line is also covered by the 2021 January 12 observing run.

\subsubsection{Shane/KAST}
\label{sec:KAST}
We observed WD\,J1653$-$1001 using the KAST Double Spectrograph on the Shane 3-m Telescope at the Lick Observatory in California, USA. We utilized the default set-up of the KAST spectrograph, using a Fairchild $2000 \times 2000$\,pixel$^2$ CCD in the blue arm and a Hamamatsu $2000 \times 4000$\,pixel$^2$ CCD in the red arm. We observed with a D57 dichroic and a 600/4310 grism for the blue side and a 830/8460 grating for the red side, with respective dispersions of 0.43\,\AA/pixel and 1.02\,\AA/pixel. The approximate wavelength range covered by the blue arm was $3600$~--~$5300$\,\AA\ and by the red arm was $5700$~--~$7800$\,\AA. We used a slit width of $1$\,arcsec, which achieved a resolution of $\approx 1$\,\AA\ in the blue arm and $\approx 2$\,\AA\ in the red arm.

The first observation took place on 2018 May 22, with one 3000\,s exposure taken in the blue arm and three consecutive 1000\,s exposures taken in the red arm. The second observation was taken on 2023 May 15, with two consecutive 2000\,s exposures taken in the blue arm and four consecutive 1000\,s exposures taken in the red arm. Spectroscopic coverage was achieved for all Balmer lines from H$\alpha$ to H$\zeta$.

\subsection{\textit{TESS} observations of \texorpdfstring{WD\,J0412$+$7549}{WDJ0412+7549}}
\label{sec:TESS}

The \textit{TESS} spacecraft observed WD\,J0412$+$7549 under designation TIC 103222871 in Sectors 19, 25 and 26 during Cycle 2, Sectors 52 and 53 in Cycle 4, and Sector 59 in Cycle 5. Observations were taken between 2019 November 28 and 2022 December 23 (see Table\,\ref{tab:tess_dates_and_periods}). Exposure times of 120\,s were taken in all six sectors. WD\,J0412$+$7549 has a \textit{TESS} magnitude of $T\simeq 15.7$.

Pre-search Data Conditioning Simple Aperture Photometry (PDCSAP) light curves were used for our analysis in Section~\ref{sec:Photometric variability of WDJ0412} as these have systematic errors removed, including error sources from the telescope and the spacecraft \citep{Stumpe2012, Smith2012}. The PDCSAP light curves for WD\,J0412$+$7549 were retrieved from the Mikulski Archive for Space Telescopes (\href{https://mast.stsci.edu/portal/Mashup/Clients/Mast/Portal.html}{MAST}) public data portal. We used the \texttt{tsa} context within \textsc{midas}\footnote{\href{https://www.eso.org/sci/software/esomidas/}{MIDAS is available from the European Southern Observatory}} to carry out the time-series analysis of the \textit{TESS} data. 

\subsection{ZTF observations of \texorpdfstring{WD\,J1653$-$1001}{WDJ1653-1001}}
\label{sec:ZTF}

ZTF is a robotic time-domain survey which uses the 48-inch Schmidt Telescope at the Palomar Observatory in California, USA \citep{Masci2019}. In this work, we use DR15 observations of WD\,J1653$-$1001 which were taken in the green- ($g$) and red- ($r$) bands of ZTF between 2018 March 17 and 2022 November 09. The light curves were retrieved from the public NASA/IPAC Infrared Science Archive (\href{https://irsa.ipac.caltech.edu/frontpage/}{IRSA}). Exposure times of 30\,s were taken in all observations.

\section{Analysis}
\label{sec:Analysis}

\subsection{Photometric and spectroscopic variability}
\label{sec:Variability}

All spectroscopic analysis of WD\,J0412+7549 in this work was performed on the IRF corrected Keck data and flux calibrated Gemini, INT and WHT data. All observation time-stamps for WD\,J0412+7549 and WD\,J1653$-$1001 were converted to a Barycentric Julian Date (BJD) Barycentric Dynamical Time (TDB). The time format used for all data is Barycentric Modified Julian Date (BMJD) minus 50\,000 which is BJD(TDB) $-$ 2\,450\,000.5.

\subsubsection{Photometric variability of \texorpdfstring{WD\,J0412$+$7549}{WDJ0412+7549}}
\label{sec:Photometric variability of WDJ0412}

\citet{Walters2021} analyzed a single sector of \textit{TESS} data of WD\,J0412$+$7549 and identified photometric variability with a period of $2.28910 \pm 0.00002$\,h. Here, we computed discrete Fourier transforms individually for all six sectors of \textit{TESS} data obtained so far (see Section~\ref{sec:TESS}). All power spectra contained a single, strong signal at a period of $\simeq2.29$\,h. We determined the spin period and its uncertainty by performing a sine-fit to the data using $\Delta \mathrm{flux} = A\mathrm{sin}(2\pi t / P - \phi) + c$ in each sector, where $A$ is the amplitude, $t$ is the observation time of each measurement, $P$ is the period, $\phi$ is the phase-shift and $c$ is an offset. The results are reported in Table\,\ref{tab:tess_dates_and_periods}. The \textit{TESS} ephemeris closest to the centre of the entire dataset which corresponds to the photometric maximum is ${\rm BMJD-50\,000} = 9368.75658(31) + 0.095379756(93)$\,\textit{E}, so we chose this as an epoch $T_\mathrm{0}$ to phase all \textit{TESS} light curves to the same baseline.

\begin{table}
\caption{Dates of the six \textit{TESS} observations of WD\,J0412$+$7549 and white dwarf spin periods measured from sine-fits to each sector independently. A combined period was measured from combining all sectors and fitting a sine function. Amplitudes from the sine-fits are reported here.}
\centering
\begin{tabular}{llcc}
\hline
Sector & Dates &  Period & Amplitude \\
 & & [h] & [per cent] \\
\hline
19 & 2019 Nov 28 -- Dec 23 & 2.28942(22) & 2.69 $\pm$ 0.13 \\
25 & 2020 May 13 -- Jun 08 & 2.28942(20) & 2.68 $\pm$ 0.12 \\
26 & 2020 Jun 08 -- Jul 04 & 2.28930(28) & 2.23 $\pm$ 0.13 \\
52 & 2022 May 18 -- Jun 13 & 2.28897(27) & 2.05 $\pm$ 0.12 \\
53 & 2022 Jun 13 -- Jul 09 & 2.28919(23) & 2.16 $\pm$ 0.11 \\
59 & 2022 Nov 26 -- Dec 23 & 2.28883(21) & 2.22 $\pm$ 0.10 \\
\hline
Combined &  & 2.2891144(16) & 2.29 $\pm$ 0.05 \\
\hline
\end{tabular}
\label{tab:tess_dates_and_periods}
\end{table}

\begin{figure*}
\centering
\includegraphics[width=2\columnwidth]{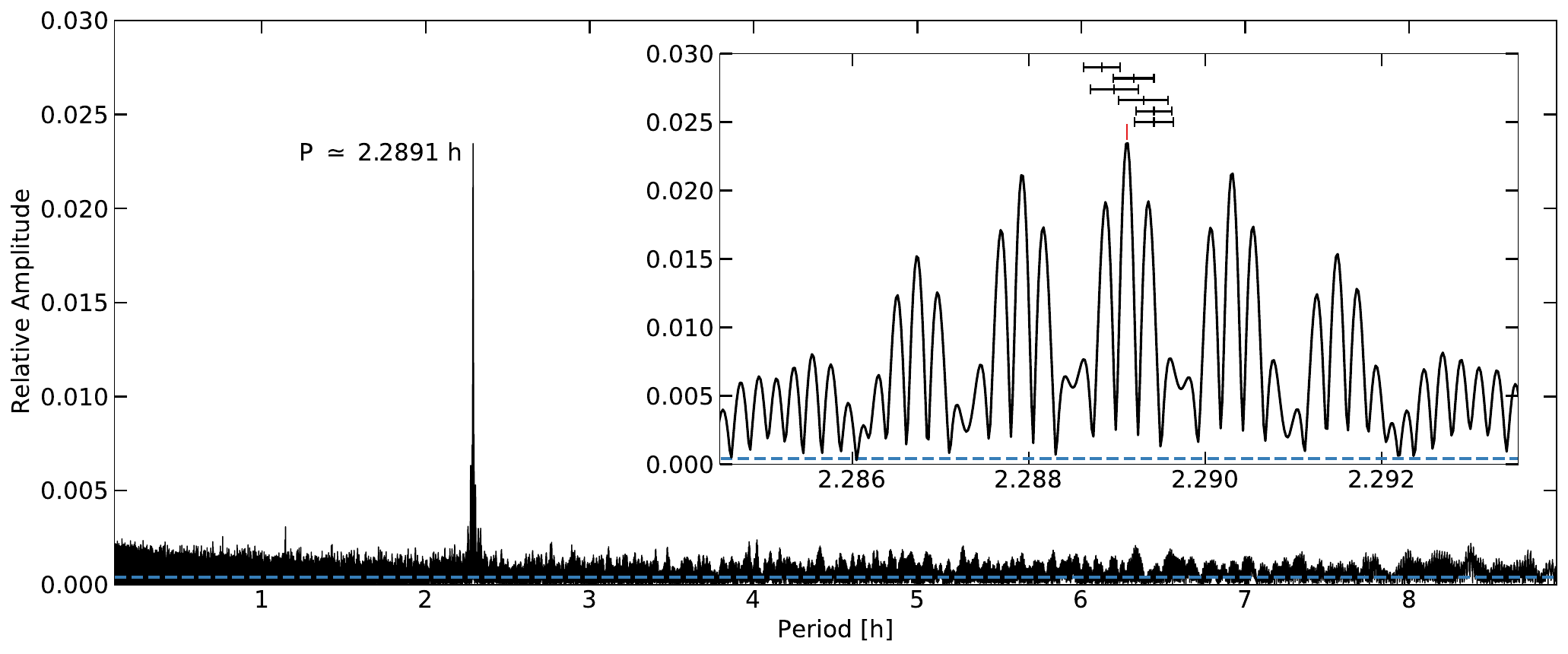}
    \caption{Power spectrum computed from the combined \textit{TESS} observations of WD\,J0412$+$7549. The strongest periodic signal is at $P \simeq 2.2891$\,h and is detected above a FAP of one per cent (blue dashed line). Inset is a zoom-in on the strongest signal, where the red tick above the central signal represents the uncertainty of the period determined from a sine fit to the combined \textit{TESS} data and the six black error bars illustrate the periods and uncertainties derived from the six individual \textit{TESS} sectors (see Table~\ref{tab:tess_dates_and_periods}).}
\label{fig:tess_power_spectrum}
\end{figure*}

Further, we computed a discrete Fourier transform of the entire \textit{TESS} observations (Figure\,\ref{fig:tess_power_spectrum}). No other strong peaks were detected across the periodogram and the period at $\simeq2.29$\,h is detected above a false alarm probability (FAP) of one per cent. The inset of Figure\,\ref{fig:tess_power_spectrum} shows the power spectrum around the confirmed period, and displays a complex alias pattern which is consistent with the window function due to the sparse sampling across the full baseline spanned by the \textit{TESS} data. 

\begin{figure}
\centering
\includegraphics[width=\columnwidth]{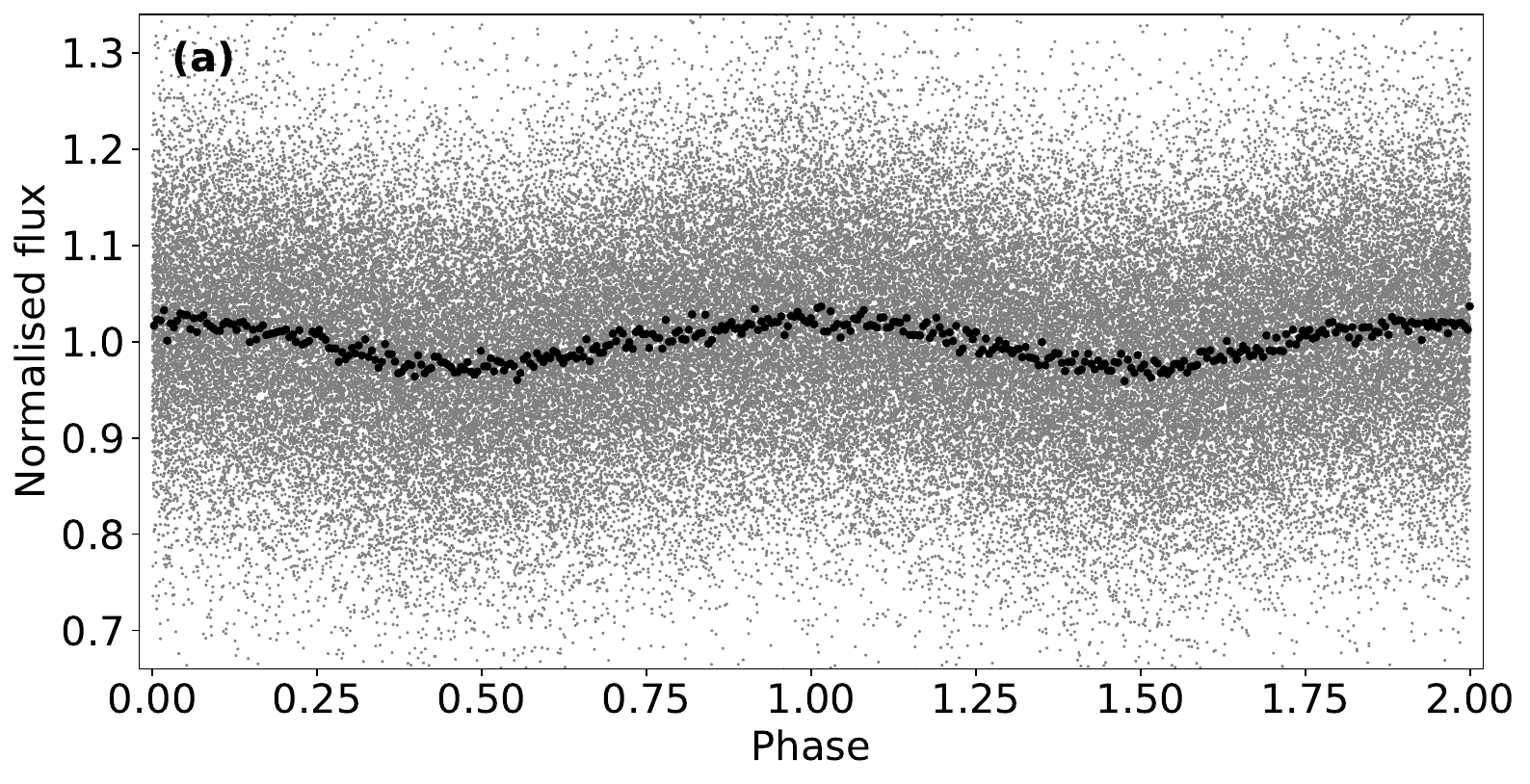}
\includegraphics[width=\columnwidth]{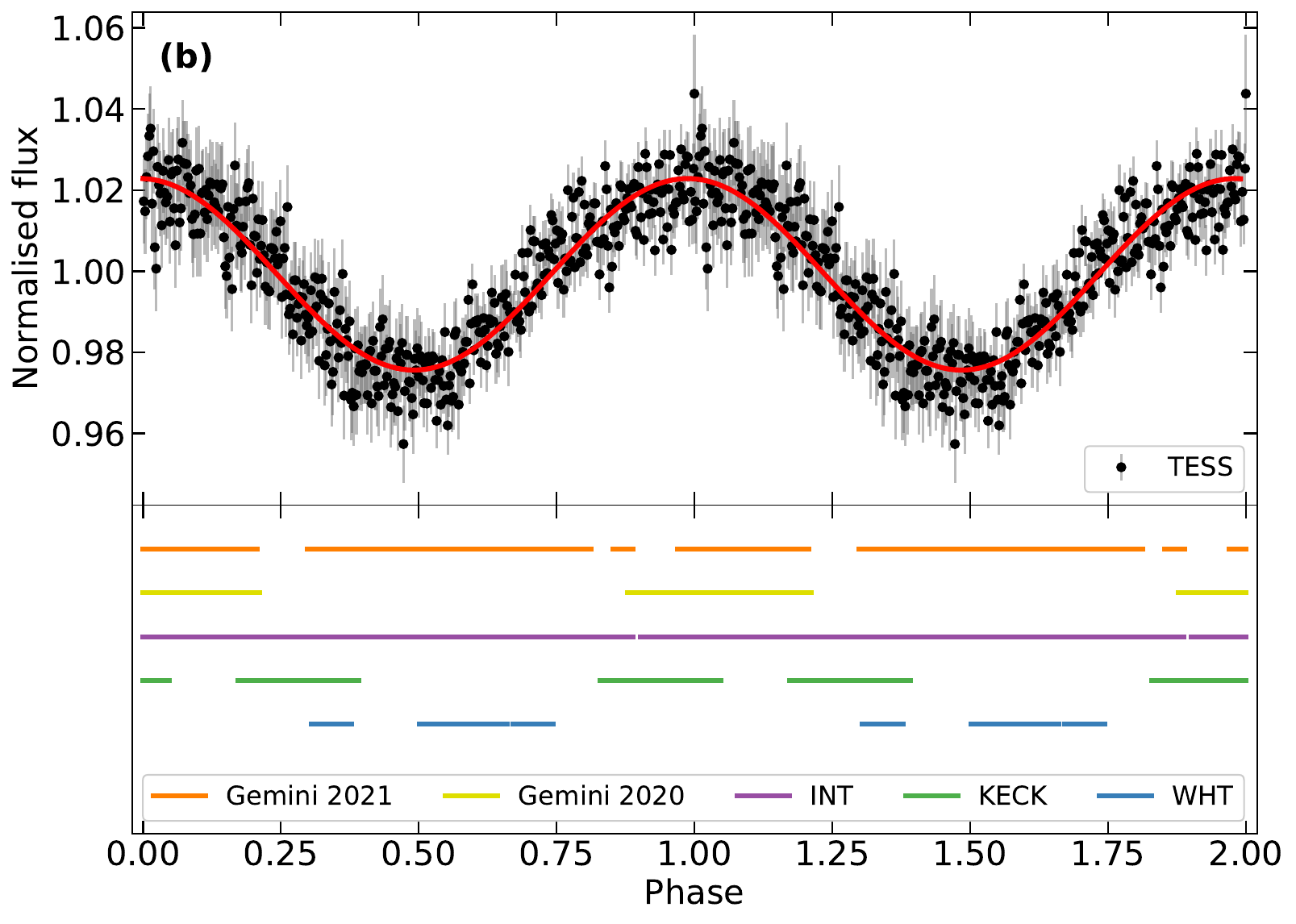}
    \caption{The combined \textit{TESS} data from the individual 120\,s cadence observations of WD\,J0412$+$7549 showing (a) all data points (grey) folded at the best-fitting period and the same data binned into 400 data points (black), and (b) the binned data points (black) fitted with a sine wave (red overlay) at the same period as the light curve and the phase of the spectroscopic observations. Full spectroscopic phase coverage by the WHT (blue), Keck (green), INT (purple) and Gemini telescopes is achieved. Gemini observations taken in 2020 (yellow) and 2021 (orange) are distinguished. The data in all panels are repeated over two phases for illustrative purposes. ${\rm Phase} = 0$ corresponds to the photometric maximum at $T_{\rm BMJD-50\,000} = 9368.75658(31)$\,d. Error bars are not shown in (a) for clarity but are shown in (b) to represent the 1$\sigma$ scatter in each bin.}
\label{fig:tess_lc_phase}
\end{figure}

A magnetic field can result in inhomogeneous brightness distributions across the white dwarf surface, which in turn leads to photometric variability on the white dwarf spin period \citep[e.g.][]{Brinkworth2013}. The combined data from the six \textit{TESS} sector observations clearly show the photometric variability of WD\,J0412$+$7549 (Figure~\ref{fig:tess_lc_phase}). The top panel of Figure~\ref{fig:tess_lc_phase} shows the light curve consisting of all data points folded upon the best-fitting period, in addition to the same data binned into 400 phase bins. A sinusoidal shape is visible when the light curve is extended over two phases. The bottom panel of Figure~\ref{fig:tess_lc_phase} shows the zoomed-in light curve on the 400 phase bins with a fitted sinusoid overlaid in red. The period obtained from a sine fit to the combined \textit{TESS} data is $2.2891144(16)$\,h. The period uncertainty is much smaller than the separation between the three central aliases in the power spectrum, and we hence conclude that this is an unambiguous measurement of the spin period of the white dwarf throughout the \textit{TESS} observations~--~under the assumption that the period is constant. 

We investigated the possibility that the period measured for WD\,J0412$+$7549 is actually half the period of variability, due to there being potentially two poles or emission spots \citep{Manser2023, Reding2023}. We inspected the phase-folded light curves on both $P = 2.2891144(16)$\,h and $2P$, but no additional structures were evident on twice the period. Therefore, we assume the periodic signal detected is likely the spin-period of the white dwarf assuming a single emission spot.

Inspecting the \textit{TESS} spin measurements (Table~\ref{tab:tess_dates_and_periods}) may suggest a trend of a decrease with time. The reduced $\chi^2$ of the periods measured from the six \textit{TESS} sectors against the assumption of a constant period is $\approx 1.5$, which hints that the period may not be constant. However, an observed minus calculated ($O-C$) analysis does not evidence a significant period change, as the line of best-fit gradient is $-0.007 \pm 0.061$ with a reduced $\chi^2 \approx 1.03$. We also performed an $\textit{F}$-test to determine the significance of a linear trend to the full set of data, testing the null-hypothesis that a linear trend is not reflective of the data presented. Following the methodology described in \citet{Munday2023}, we find an $\textit{F}$-ratio of 20.53 which, under the $\textit{F}$(1, $n-2$) distribution for our $n=6$ measurements, indicates that there is a 98.94 per cent (2.6$\sigma$) significance of a linear trend in the data. We note that the apparent decreasing trend in the individual periods may be related to small phase or period drifts, or simply be an artifact of the instrument. For now, we caution against over-interpreting the apparent trend and recommend that the spin period of WD\,J0412+7549 should keep being monitored.

In the bottom panel of Figure~\ref{fig:tess_lc_phase}, the phase coverage of each exposure taken with the WHT, Keck, INT and Gemini telescopes are represented with horizontal coloured bars. WD\,J0412$+$7549 was observed on three consecutive nights with the WHT, which covered $\approx 31$ per cent of the periodic signal identified. The two 30-minute exposures taken by Keck resulted in $\approx 44$ per cent of the spin period being spectroscopically covered. The INT exposures taken over two consecutive nights covered $\approx 98$ per cent of the spin period. The Gemini exposures taken in 2021 covered $\approx 79$ per cent of the spin period, but together with the 2020 exposures $\approx 87$ per cent was covered. Considering all of the above observations, we have achieved full spectroscopic phase coverage of WD\,J0412$+$7549 with short-cadence observations which allowed us to analyse the emission line variability over the entire spin period (Section~\ref{sec:Spectroscopic variability}).

\subsubsection{Photometric variability of \texorpdfstring{WD\,J1653$-$1001}{WDJ1653-1001}}
\label{sec:Photometric variability of WDJ1653}

\begin{figure*}
\centering
\includegraphics[width=1.39\columnwidth]{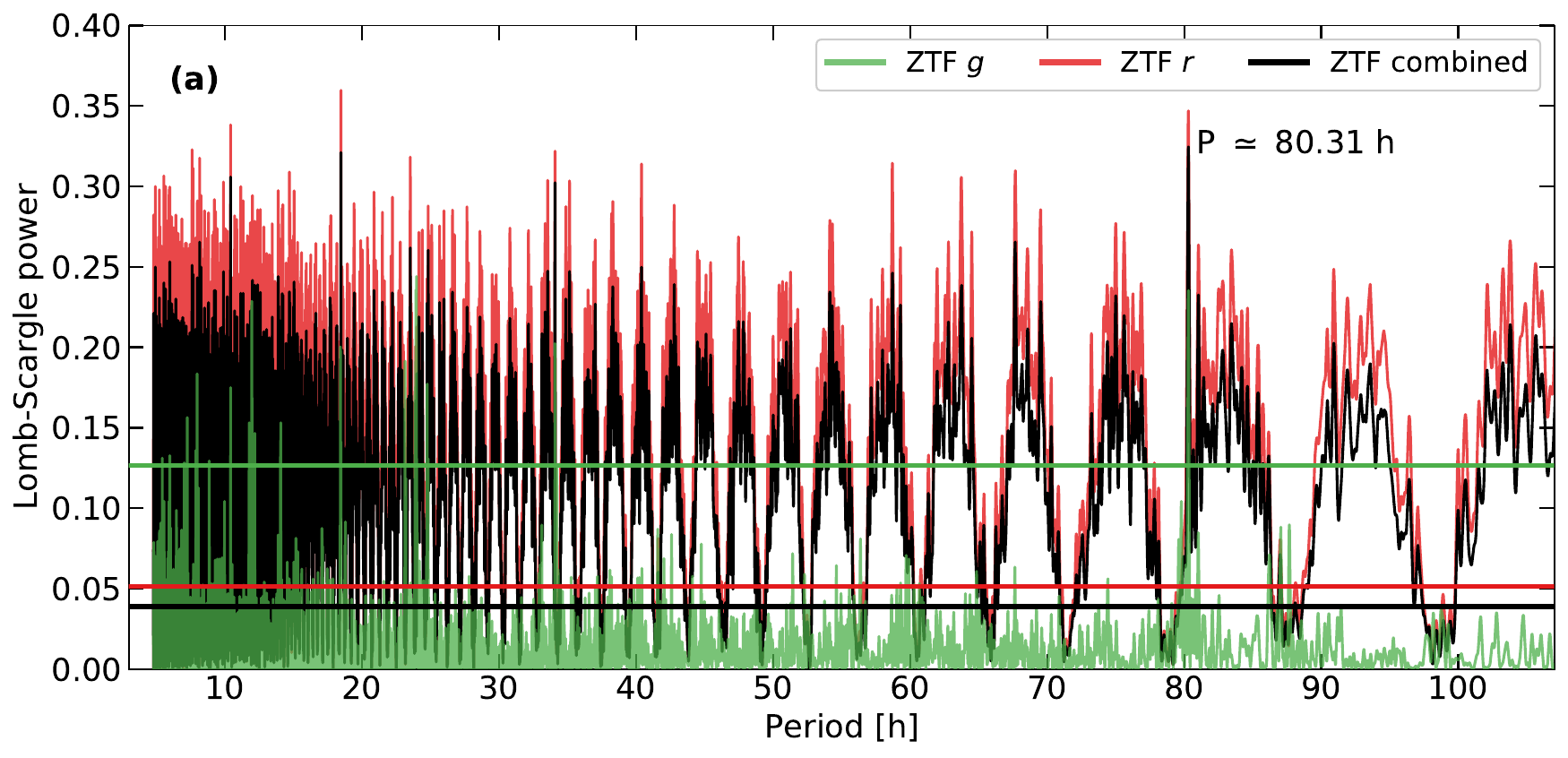}
\includegraphics[width=0.67\columnwidth]{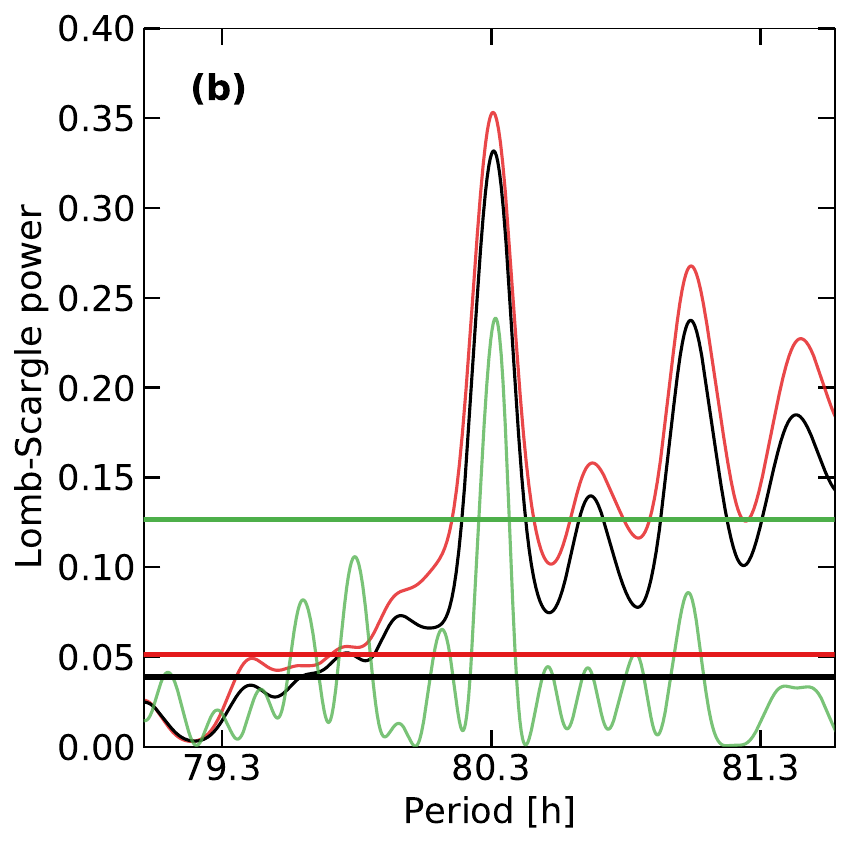}
    \caption{Power spectra computed from the $g$-band, $r$-band and combined $g$- and $r$-band DR15 ZTF data of WD\,J1653$-$1001. The signal common to all power spectra out of the five strongest signals in the individual power spectra corresponds to a period of $\simeq 80.31$\,h, which is detected above a FAP of one per cent (green, red and black solid lines). Panel (b) shows the power spectra zoomed in on the periodic signal at $\simeq 80.31$\,h. The legend applies to both panels.}
\label{fig:ZTF_power_spectra}
\end{figure*}

The $g$-band and $r$-band flux of WD\,J1653$-$1001 was calculated from the ZTF magnitude data, relative to the median magnitude in each band. We combined the $g$-band and $r$-band datasets and weighted the contribution of the individual band points equally. Then, we computed Lomb-Scargle periodograms \citep{Lomb1976, Scargle1982} using the \texttt{python} package \texttt{astropy.timeseries} \citep{Astropy2013, Astropy2018, Astropy2022} for the individual $g$-band and $r$-band light curves in addition to the combined $g$- and $r$-band light curve (Figure~\ref{fig:ZTF_power_spectra}). 

The strongest signals in all three power spectra occur at periods $\gtrsim 11.5$\,h. ZTF is sensitive to shorter periods than \textit{TESS} due to shorter exposure times, so if WD\,J1653$-$1001 had a similar period with similar amplitude to that found for WD\,J0412+7549 then it would be detected. The strongest signal in the $g$-band power spectrum is at a period of $\simeq 29.4729644$\,h which we attribute to the moon and not as a true measurement of the white dwarf spin period. 

Out of the five strongest signals in all three power spectra, only the periodic signal at $\simeq 80.31$\,h is common to all three and is detected above a FAP of one per cent\footnote{The modulation in the $g$- and $r$-band are near identical so the FAP of the combined datasets is a very good approximation.}. The period of $\simeq 80.31$\,h is the strongest signal in the combined $g$- and $r$-band power spectrum, the second strongest signal in the $r$-band power spectrum and the third strongest signal in the $g$-band power spectrum. However, with the current ZTF data it is not possible to unambiguously determine whether the peaks at $\simeq 80.31$\,h are aliases related to the sampling rate or are indeed the intrinsic periodic signal. We note that the periodic signal at $\simeq 18.48$\,h is the strongest signal in the $r$-band and second strongest signal in the combined $g$- and $r$-band power spectra, however it is the 23rd strongest signal in the $g$-band power spectrum, so is not considered the dominant periodic signal at this time.

\begin{figure}
\centering
\includegraphics[width=\columnwidth]{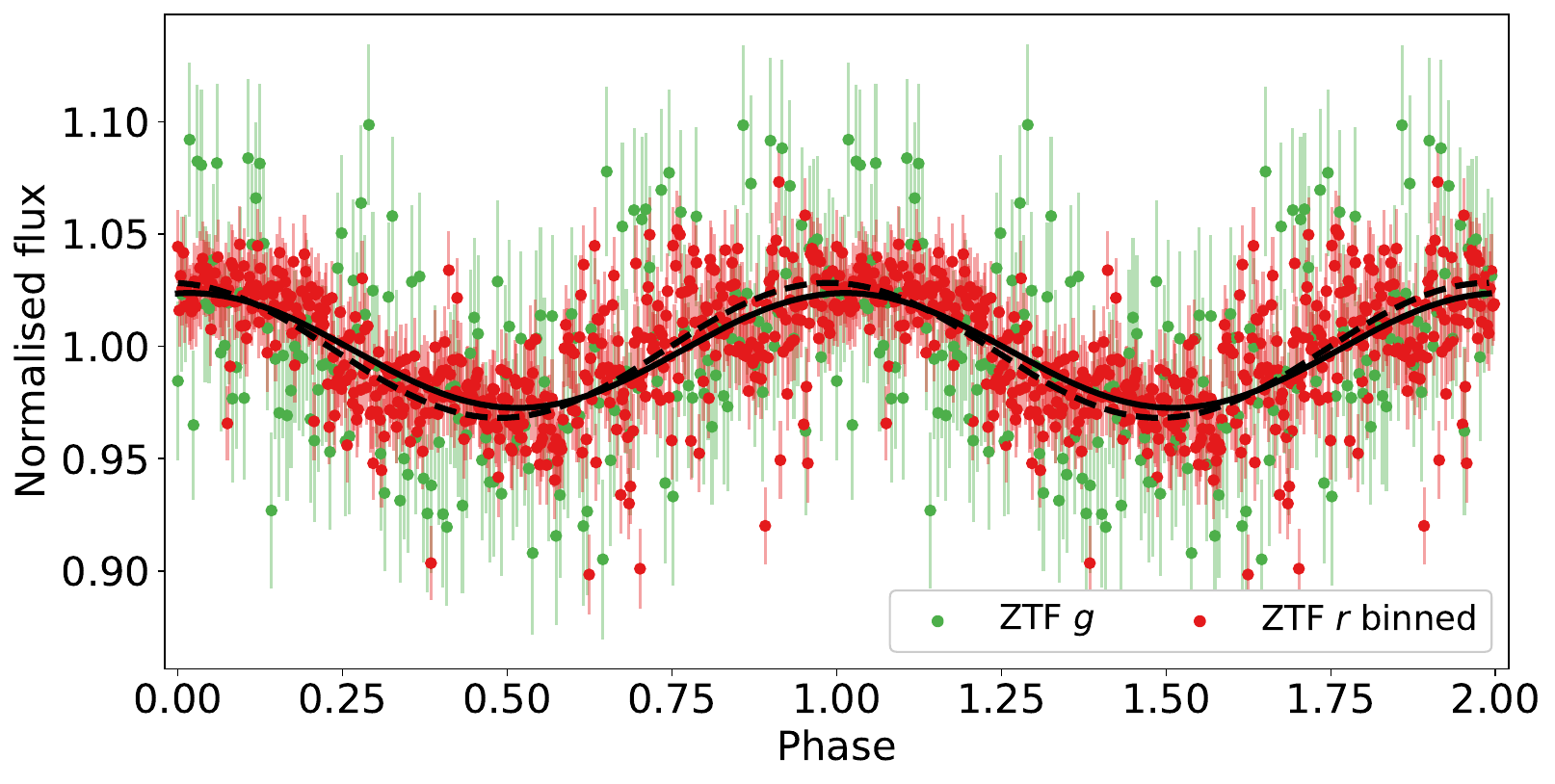}
    \caption{ZTF DR15 $g$- (green) and $r$- (red) band light curves of WD\,J1653$-$1001, phase-folded onto a period of $80.31$\,h. Sine waves were fitted on the $g$- and $r$-band light curves, shown by the black dashed and solid curves, respectively. The $g$-band raw data are shown, whereas the $r$-band data has been binned into 400 data points. The data are repeated over two phases for illustrative purposes. ${\rm Phase} = 0$ corresponds to the photometric maximum at $T_{\rm BMJD-50\,000} = 9701.44(5)$\,d for the $g$-band light curve and $T_{\rm BMJD-50\,000} = 9702.07(5)$\,d for the $r$-band light curve. Error bars are shown to represent the 1$\sigma$ scatter in each measurement/bin.}
\label{fig:ZTF_lc_phase}
\end{figure}

Despite the uncertainty related to the period, we created phase-folded light curves on a $80.31$\,h period with the $g$-band, $r$-band and combined $g$- and $r$-band data and fit them with a sinusoidal function. There is very little variation between the $g$-band and $r$-band light curves (Figure~\ref{fig:ZTF_lc_phase}), with the amplitudes of both differing by $< 1\sigma$. Therefore, we used the combined $g$- and $r$-band light curve to measure the period of WD\,J1653$-$1001 from the sine fit, which resulted in $80.534 \pm 0.087$\,h and an amplitude $2.3 \pm 0.2$ per cent. The period uncertainty is within $1\sigma$ of the strongest detected periodic signal and is much smaller than the separation between adjacent aliases in the power spectrum. Previous studies from \citet{Reding2020} and \citet{Manser2023} found a colour dependence on the strength of variability in DAHe stars, so WD\,J1653$-$1001 differs from DAHe stars in this way as no colour dependence is evident. 

We performed further analysis to determine the likelihood that the measured periodic signal of $80.534 \pm 0.087$\,h is intrinsic to WD\,J1653$-$1001, due there being a background main-sequence star $\approx 1$\,arcsec away from this white dwarf. The proper motion of WD\,J1653$-$1001 reveals that over the span of ZTF observations included in DR15, the white dwarf moved from one side of the main-sequence star to the other while almost directly passing in-front of it. There is no indication that WD\,J1653$-$1001 and the contaminant were resolved into separate sources by ZTF so we cannot exclusively determine with ZTF whether the period of variability is sourced from WD\,J1653$-$1001 or the contaminant.

To help determine whether the periodic signal is coming from WD\,J1653$-$1001 or the contaminant, we created a variability metric to investigate the likelihood of the objects being variable. Our variability metric takes a similar approach to \citet{Guidry2021}, which was used by \citet{Reding2023} to identify the DAHe stars LP\,705$-$64 and WD\,J143019.29$-$562358.33. Our metric compensates for some systematic effects and is easy to interpret: it returns a sigma value relative to the median photometric scatter of sources of similar brightness and sky position. For example, the two stars just mentioned return 3.9$\sigma$ and 5.5$\sigma$, respectively, i.e. they show significantly more photometric scatter than similar sources.

The variability metric is calculated using the scatter in individual \textit{Gaia~G}-band observations. That quantity is not directly available, so it is estimated for every source in the catalogue as  
$S=\sqrt{\textsc{phot\_g\_n\_obs}/\textsc{phot\_g\_mean\_flux\_over\_error}}$. The distribution of $S$ for sources of the same brightness should be approximately Gaussian, with a long tail of high values for variable sources. To model that distribution across the catalogue, the 16\textsuperscript{th}, 50\textsuperscript{th}, and 84\textsuperscript{th} quantiles of $S$ (denoted e.g. $S_{50}$) are determined within many magnitude bins. The local distribution of $S$ for any given source can then be estimated by interpolation. The variability of that source is then calculated as $\sigma_S=(S-S_{50})/S_{16}$ where $S<S_{50}$, or $\sigma_S=(S-S_{50})/S_{84}$ where $S>S_{50}$. However, inspection of $\sigma_S$ values across the sky of such values reveals the clear imprint of the \textit{Gaia} scanning law. Fractional residuals $S/S_{50}$ are calculated for every source in the catalogue, and the median of those residuals calculated within every level-7 HEALpix pixel. Those positional medians provide a correctional scaling factor, as the median residual within each HEALpix pixel should be unity in the absence of systematics. The scaling factor is interpolated between HEALpix pixel centres and used to correct $S$ across the entire catalogue. Quantiles of $S$ are then recalculated within each magnitude bin, and the whole procedure iterated until the positional corrections converge on unity.

We assessed photometric variability $S$ for both objects compared to a random sample of 10\,000 main-sequence stars from the same region of the \textit{Gaia} HRD as the background object. Both WD\,J1653$-$1001 and the contaminant are 5$\sigma$ outliers, meaning either both or one of them is photometrically variable but \textit{Gaia} cannot fully resolve the two objects. Of the 10\,000 main-sequence stars, 2.5 per cent are variable at or above 5$\sigma$. Using the same variability metric on the sample of DA(H)e stars revealed that the two DAe stars and one DAHe star are variable at or above 5$\sigma$. The DAHe stars have a median variability of 0.6$\sigma$, with the maximum being 5.5$\sigma$. Therefore, 10.7 per cent of DA(H)e stars and 100 per cent of DAe stars are variable above 5$\sigma$ which are both larger than the probability that the variability is coming from the contaminant main-sequence star. 

As a last check, we looked in \textit{Gaia} DR3 which has entries for WD\,J1653$-$1001 and the contaminant and found the \texttt{phot\_variable\_flag} parameter \citep{Eyer2017, Eyer2022} identified variability in the photometric data for WD\,J1653$-$1001, but unfortunately the photometric data of the main-sequence star was not processed or exported to the catalogue. The four DA(H)e stars which have a median photometric error above 3.9$\sigma$ are classified as variable in \textit{Gaia} DR3 which provides validation to our own variability metric.

The possibility of WD\,J1653$-$1001 having surface features more complex than a single spot/region beneath the chromosphere cannot be ruled out with the current ZTF data. The phase-folded light curve on $2P = 161.068$\,h has two maxima and minima per cycle which could suggest the presence of two emission spots/regions \citep{Manser2023, Reding2023}. If this is the case, then the period of variability would be $2P$, with the measured period of $80.534 \pm 0.087$\,h being an alias. However, we cannot confidently conclude either way due to the uncertainty surrounding the ZTF periodic signal analysis.

The combination of the ZTF power spectra, phase-folded light curve, \textit{Gaia} DR3 variability flag and our own variability check makes us confident that WD\,J1653$-$1001 is photometrically variable, with a dominant periodic signal of $80.534 \pm 0.087$\,h. However, only by obtaining full phase coverage of this star with dedicated follow-up time-domain observations~--~which has ideally been resolved for WD\,J1653$-$1001 and the contaminant main-sequence star~--~will it be possible to unambiguously confirm whether the measured periodic signal is the true spin period for WD\,J1653$-$1001, if it is an alias, or if it is indeed the period of variability for the contaminant.

\subsubsection{Spectroscopic variability}
\label{sec:Spectroscopic variability}

\begin{figure*}
	\includegraphics[width=\columnwidth]{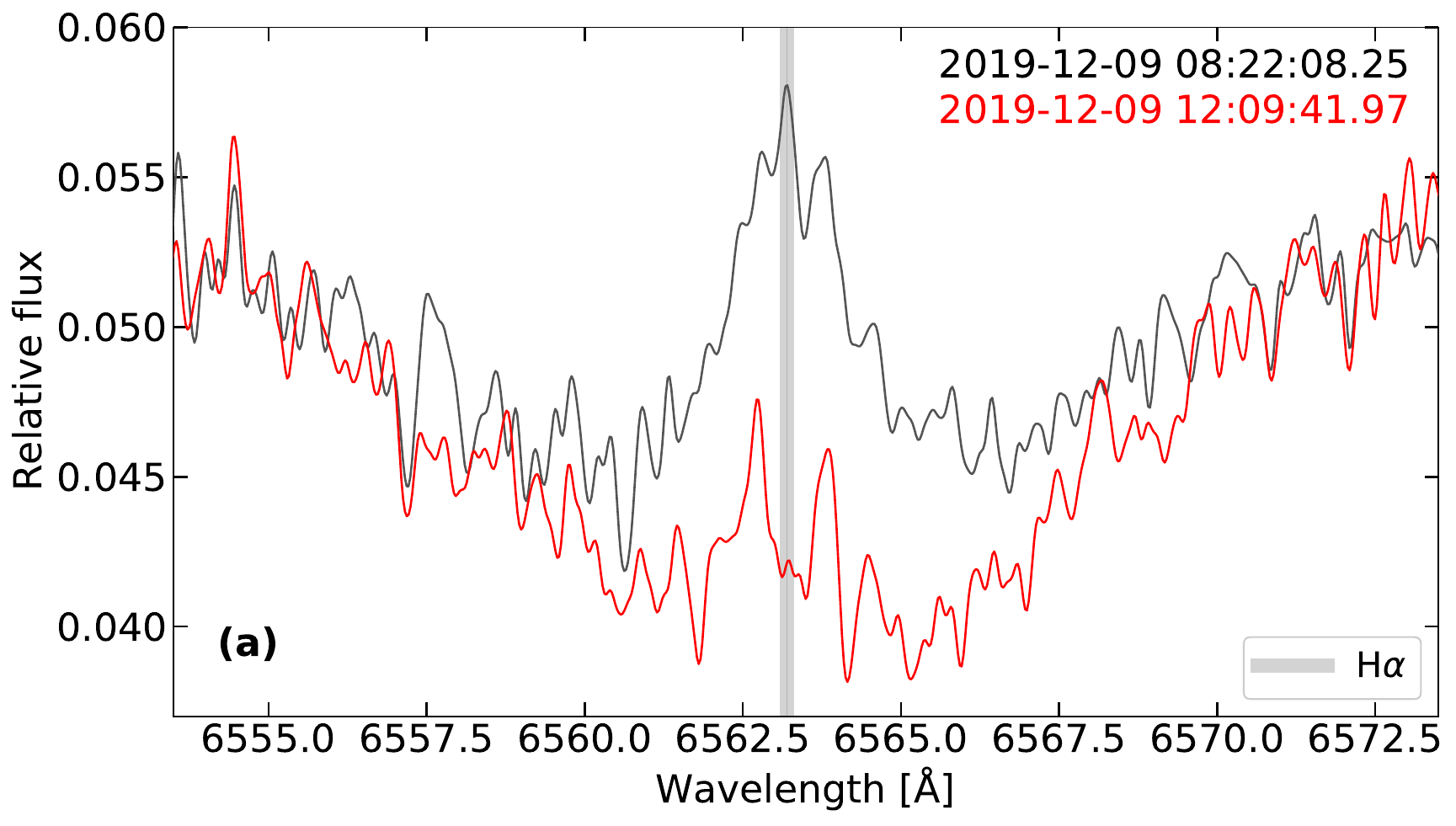}
	\includegraphics[width=\columnwidth]{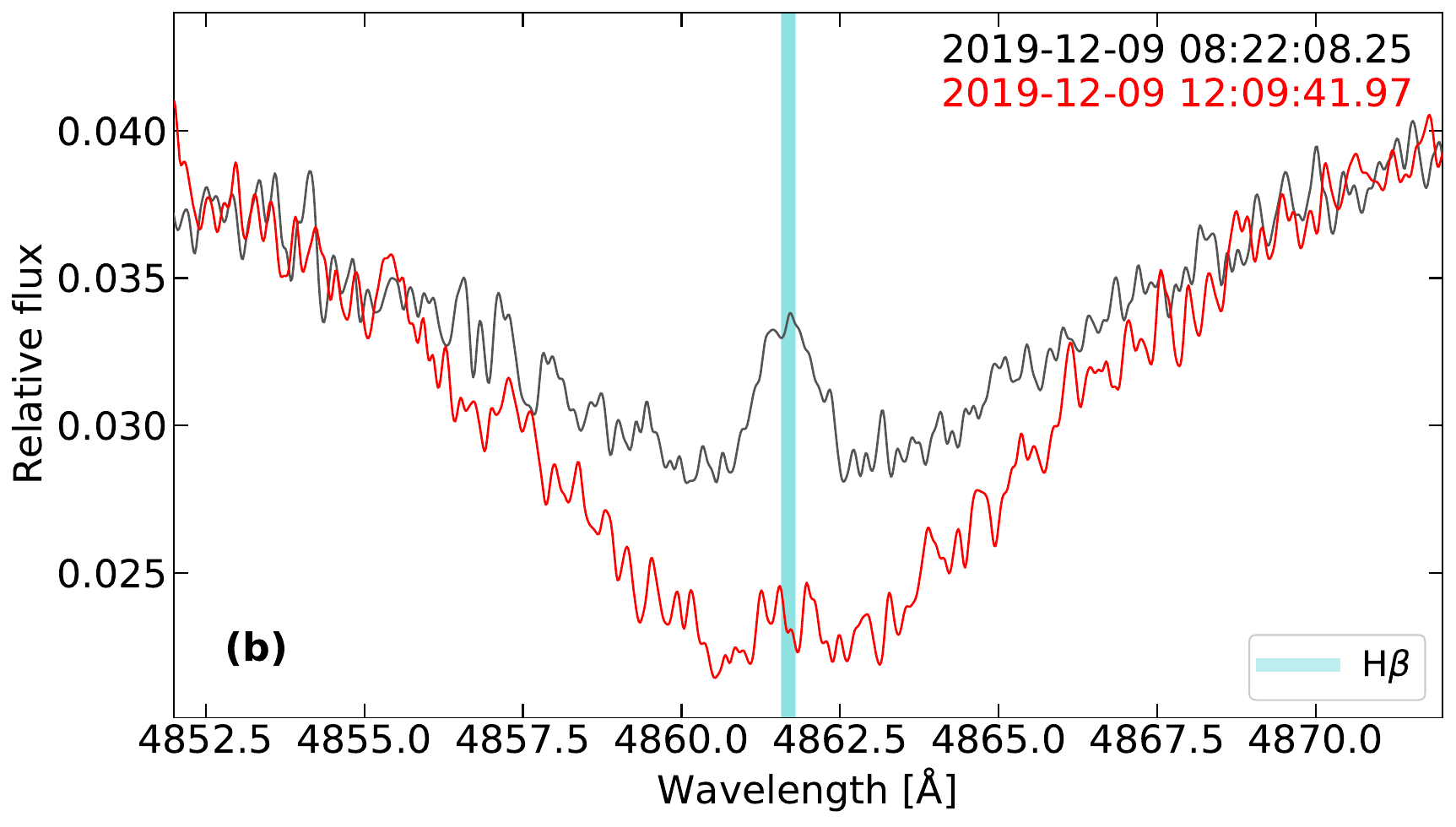}
    \caption{Two 30-minute exposures of WD\,J0412$+$7549 taken with the Keck HIRES instrument. Spectra were taken 3.79\,h apart around the emission core of (a) H$\alpha$ and (b) H$\beta$. The coloured vertical bars indicate the rest wavelengths of H$\alpha$ (grey) and H$\beta$ (aqua) corrected for radial velocity (Section~\ref{sec:radial velocity}). All spectra have had the instrumental response function removed (see text for details) and are smoothed with a 5-pixel boxcar for clarity. The observation UT date and start times are shown in the top right corner of the plots.}
    \label{fig:Keck_Ha_Hb}
\end{figure*}

Spectra taken of WD\,J0412$+$7549 using the WHT (Figure~\ref{fig:WHT_Ha_Hb_Hg_Hd_Hz}), Keck (Figure~\ref{fig:Keck_Ha_Hb}), INT (Figure~\ref{fig:INT_Ha_Hb}) and Gemini (Figure~\ref{fig:gemini_Ha_Hb_Hg_Hd}) telescopes all evidence H$\alpha$ and H$\beta$ line emission. The emission line strength clearly changes between exposures taken over the spin period, which is strong evidence of emission line variability.

WD\,J0412$+$7549 was observed by the WHT on three consecutive nights. Multiple observations were made on the first night so the stacked spectrum is presented in Figure~\ref{fig:WHT_Ha_Hb_Hg_Hd_Hz}, in addition to the single spectra taken on the subsequent two nights. Erroneous wavelength shifts were apparent in the original spectra suggesting that improper wavelength calibration had been performed. A reliable wavelength calibration could not be performed on these spectra, hence we do not pursue this avenue further.

Two 30-minute exposures covering the H$\alpha$ and H$\beta$ line regions were taken with Keck of WD\,J0412$+$7549 and are shown in Figure~\ref{fig:Keck_Ha_Hb}. It is clear that both Balmer emission lines change in strength between exposures, which were taken 227.56\,min apart. Broad emission wings are evident in both exposures, and likely result from the different temperature and pressure stratifications of the atmosphere and chromosphere in (possibly magnetic) spots. However, further investigation is required to confirm this. 

The H$\alpha$ and H$\beta$ emission line variability is seen in all exposures of the INT (Figure~\ref{fig:INT_Ha_Hb}) and Gemini (Figure~\ref{fig:gemini_Ha_Hb_Hg_Hd}) observations, where the line strength variation is observed more gradually due to shorter exposures taken almost over the entire spin period (Figure~\ref{fig:tess_lc_phase}). Only the H$\alpha$ and H$\beta$ line regions are shown in Figure~\ref{fig:INT_Ha_Hb} as the H$\gamma$ line region is cut off by the INT flux calibration and is therefore not reliable. Out of the observations performed with the Gemini telescope, the H$\alpha$ and H$\beta$ emission line variability is most clearly seen in the exposures taken on 2021 January 12 therefore we only show these spectra in Figure~\ref{fig:gemini_Ha_Hb_Hg_Hd}. Almost pure atmospheric H$\alpha$ emission and absorption features are visible in the exposures taken at 07:27:29 and 08:26:58, respectively, which occur nearly half a spin period apart (0.433 phase difference). 

\begin{figure*}
	\includegraphics[width=2\columnwidth]{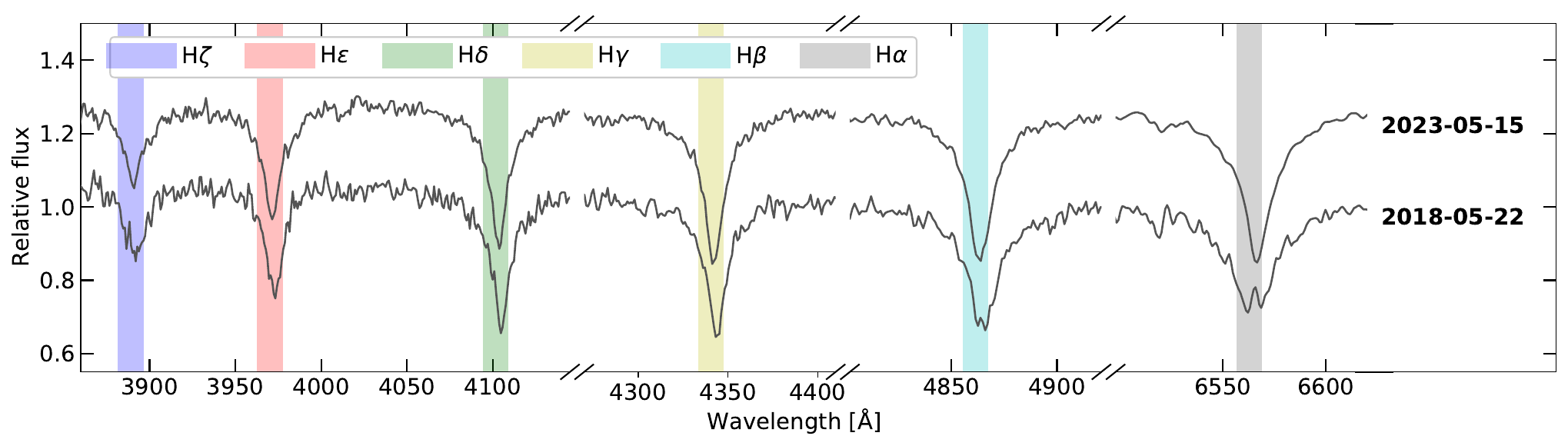}
    \caption{Spectra of WD\,J1653$-$1001 taken with the KAST spectrograph on 2018 May 22 and 2023 May 15 around the H$\alpha$ to H$\zeta$ Balmer line regions. Shown is the individual spectrum taken in the blue arm on 2018 May 22 and the stacked spectra from the other exposures. The observation UT dates are shown on the right of the plot. Spectra are convolved with a Gaussian with a FWHM of 2\,\AA\ and offset vertically for clarity.}
    \label{fig:KAST_Ha_Hb_Hg_Hd_Hz}
\end{figure*}

The spectra of WD\,J1653$-$1001 taken with the KAST instrument on 2018 May 22 and 2023 May 15 are shown in Figure~\ref{fig:KAST_Ha_Hb_Hg_Hd_Hz}, which consists of the individual exposure taken in the blue arm on 2018 May 22 and the stacked spectra from the other exposures. Coverage of the H$\alpha$ to H$\zeta$ line regions was achieved, where the emission cores of H$\alpha$ to H$\beta$ are clearly visible.

None of the Balmer emission lines in WD\,J0412$+$7549 nor WD\,J1653$-$1001 spectra exhibit Zeeman-splitting therefore no magnetic field detection or strength measurements could be made. Instead, we measured the upper limit of the magnetic field strength using the first Keck exposure of WD\,J0412$+$7549, as it has the highest resolution out of all the observations, and the first KAST spectrum of WD\,J1653$-$1001. We constructed a delta function and convolved it with the spectral resolution of the instrument. We then Zeeman-split the delta function, overlaid it on the H$\alpha$ emission core and altered the magnetic field strength until the delta function was wide enough that we would visually see Zeeman-splitting and be able to distinguish it from the noise. This technique yielded upper limits on the magnetic field strength $B < 0.05$\,MG for both DAe stars. The magnetic field limit obtained from Keck is limited by the intrinsic widths of emission features and the length of the exposures (0.219 of the phase). We cannot exclude the possibility of spin-related magnetic smearing in the emission core over the exposure, or intrinsic broadening from a high temperature chromosphere. Hence, we obtain a similar magnetic field limit from Keck and KAST for both white dwarfs, despite the difference in instrumental resolution.

\begin{figure*}
	\includegraphics[width=\columnwidth]{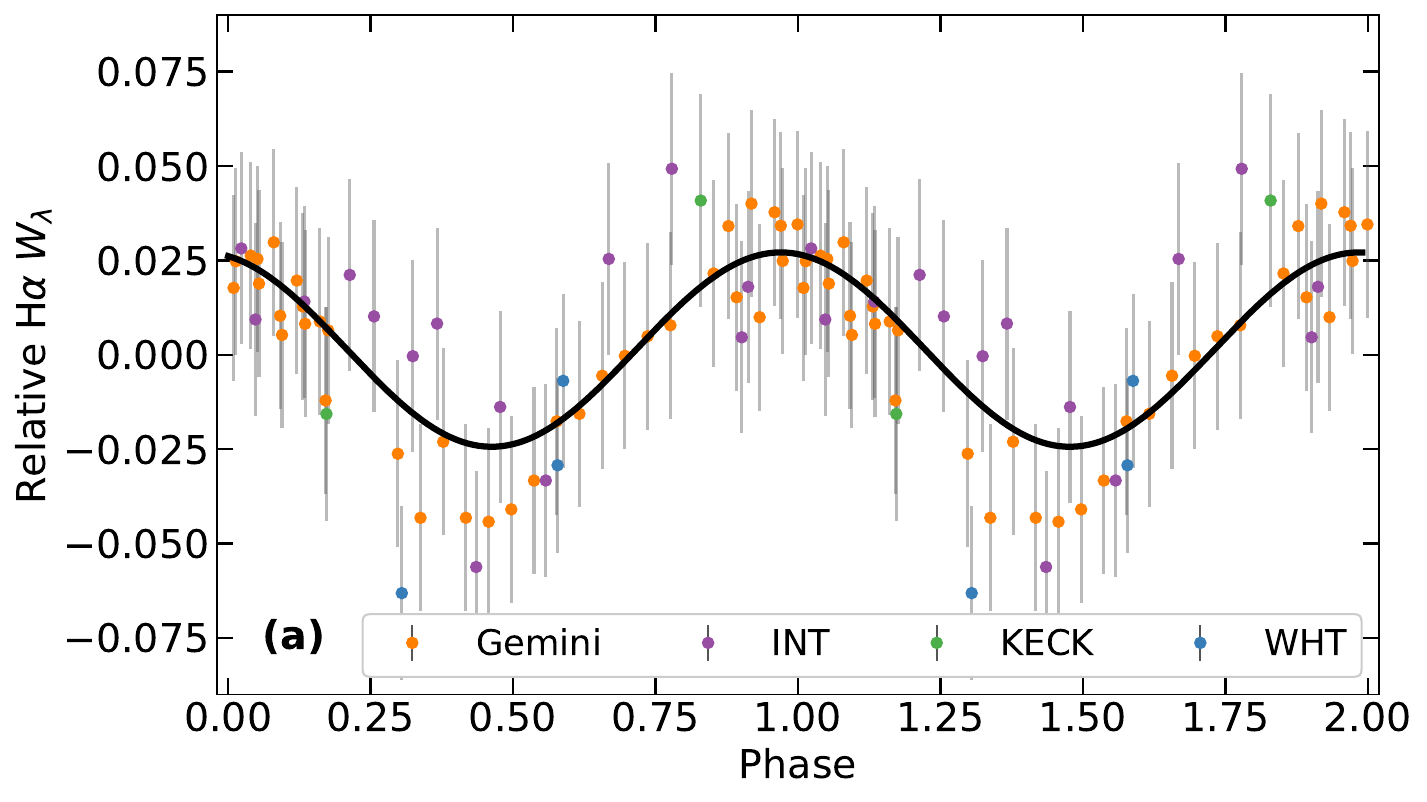}
	\includegraphics[width=\columnwidth]{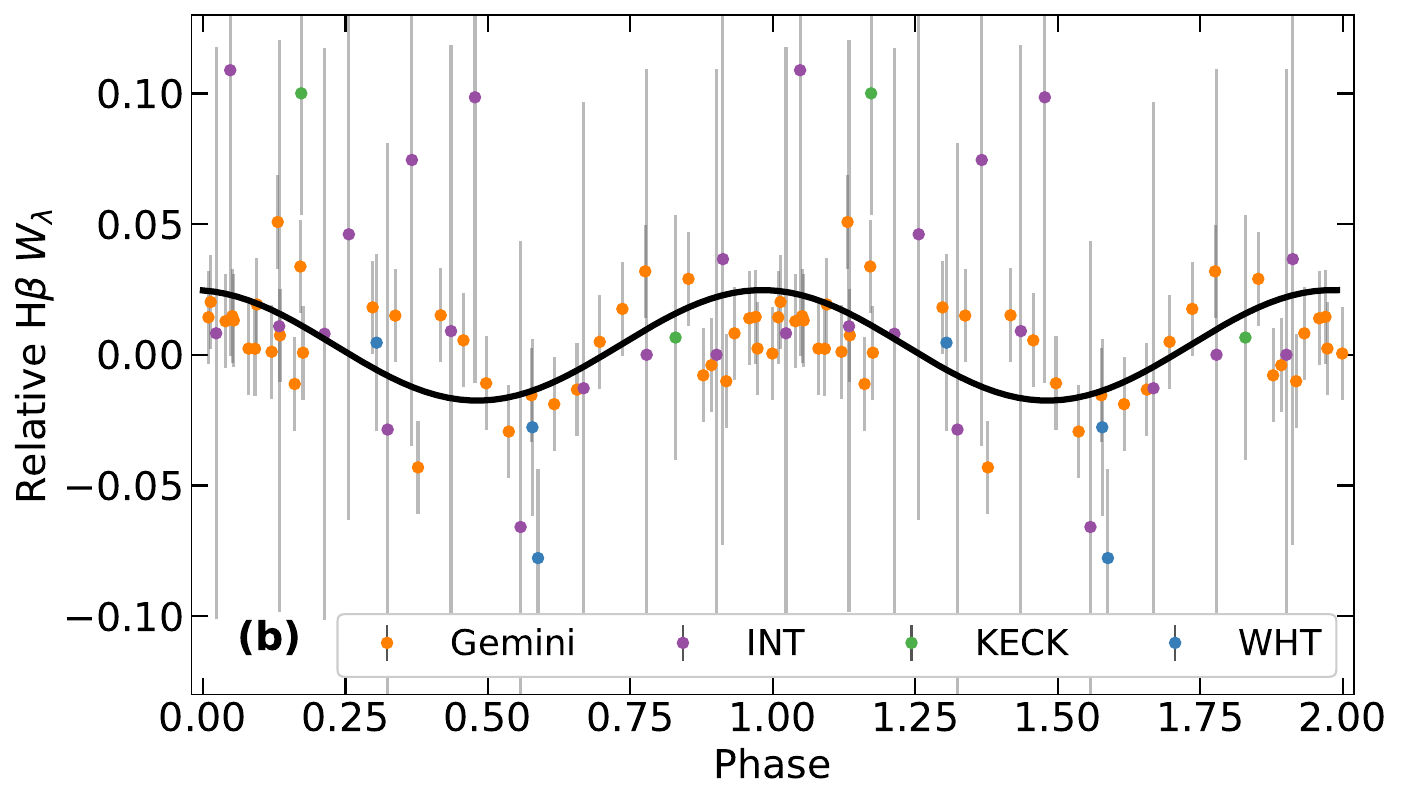}
    \caption{The phase of WD\,J0412$+$7549 as a function of the relative equivalent width (\EW) of its spectroscopically observed (a) H$\alpha$ and (b) H$\beta$ Balmer lines by the WHT (blue), Keck (green), INT (purple) and Gemini (orange) telescopes. The data are fitted with a sine wave (black overlay) and repeated over two phases for illustrative purposes. Weakest emission corresponds to the largest \EW, which occurs at $\phi = 0$ (i.e. photometric maximum). Error bars are correspond to 1$\sigma$ uncertainties.}
    \label{fig:phase_EW_Ha_Hb}
\end{figure*}

To investigate the emission activity of WD\,J0412$+$7549, we measured the equivalent widths (\EW) of the H$\alpha$ and H$\beta$ emission cores in all exposures, where the smaller the value of \EW\ corresponds to stronger emission. The \EW\ were normalized using the \texttt{python} package \texttt{scikit-learn} \citep{scikit-learn} with an L2 normalization, then compared against the phase of each exposure. On average, the \EW\ uncertainties are larger for H$\beta$ emission cores which are noisier and shallower than the H$\alpha$ emission cores. Figure~\ref{fig:phase_EW_Ha_Hb} displays the \EW\ of the H$\alpha$ and H$\beta$ emission line cores as a function of phase, where $\phi = 0$ corresponds to the photometric maximum at $T_{\rm BMJD-50\,000} = 9368.75658(31)$\,d. The data are repeated over two phases for clarity. We fitted a sinusoid to the data to clearly show that the weakest emission (i.e. largest \EW) occurs at $\phi = 0$. In contrast, Figure~\ref{fig:tess_lc_phase} shows that the photometric flux maximum occurs at $\phi = 0$. Hence, there is an anti-phase relationship between the photometric and emission line variability of WD\,J0412$+$7549. 

The same analysis of \EW\ against phase could not be performed for WD\,J1653$-$1001 as there is not enough time-resolved spectroscopic data for this star. However, we measured the \EW\ of the H$\alpha$ emission feature in the three KAST exposures taken in the red arm on 2018 May 22, and found that all three have an identical \EW. The lack of spectral variability over the 3000\,s baseline would be consistent with a large period, possibly with the one found by ZTF (Section~\ref{sec:Photometric variability of WDJ1653}), and the spectra taken on 2023 May 15 (Figure~\ref{fig:KAST_Ha_Hb_Hg_Hd_Hz}) confirms variability over a longer baseline.

\subsection{Photometric and spectroscopic parameters}
\label{sec:parameters}

Photometric and spectroscopic fits were performed to calculate the atmospheric parameters of WD\,J0412$+$7549 and WD\,J1653$-$1001. The best-fitting model for DAe stars was ascertained by testing three DA model atmospheres: 1D radiative, 1D convective and 3D convective \citep{Tremblay2013,Tremblay2015}. 

The photometric fits of WD\,J0412$+$7549 and WD\,J1653$-$1001 were performed using photometry from \textit{Gaia} DR3, Pan-STARRS and 2MASS (Table~\ref{tab:photometry}). Both DAe stars have $\approx 2.3$ per cent flux variability (see Sections~\ref{sec:Photometric variability of WDJ0412}~and~\ref{sec:Photometric variability of WDJ1653}) so, based on Eq.16 from \citet{GF2019}, we imposed a lower limit of 0.025\,mag on the photometric uncertainties from Pan-STARRS and 2MASS. There are 395 and 599 repeat observations of WD\,J0412$+$7549 and WD\,J1653$-$1001 in \textit{Gaia} DR3, respectively, thus the $G$, $G_{\rm BP}$ and $G_{\rm RP}$ measurements stated in \textit{Gaia} DR3 are likely averages over each star's full rotation period which are suitable to use to calculate their photometric parameters. The \textit{Gaia} DR3 measured uncertainties are likely more precise than our imposed lower limit on the Pan-STARRS and 2MASS measurements, therefore we did not modify these uncertainties.

We tested the photometric fits using the three DA model atmospheres however the differences between atmospheric parameters were negligible. The best-fitting atmospheric parameters were found to be $\Teff = 8546 \pm 87$\,K and $\log g = 8.260 \pm 0.030$\,dex for WD\,J0412$+$7549 and $\Teff = 7388 \pm 71$\,K and $\log g = 7.930 \pm 0.030$\,dex for WD\,J1653$-$1001 (Table~\ref{tab:stellar parameters}).

\begin{figure}
\centering
\includegraphics[width=\columnwidth]{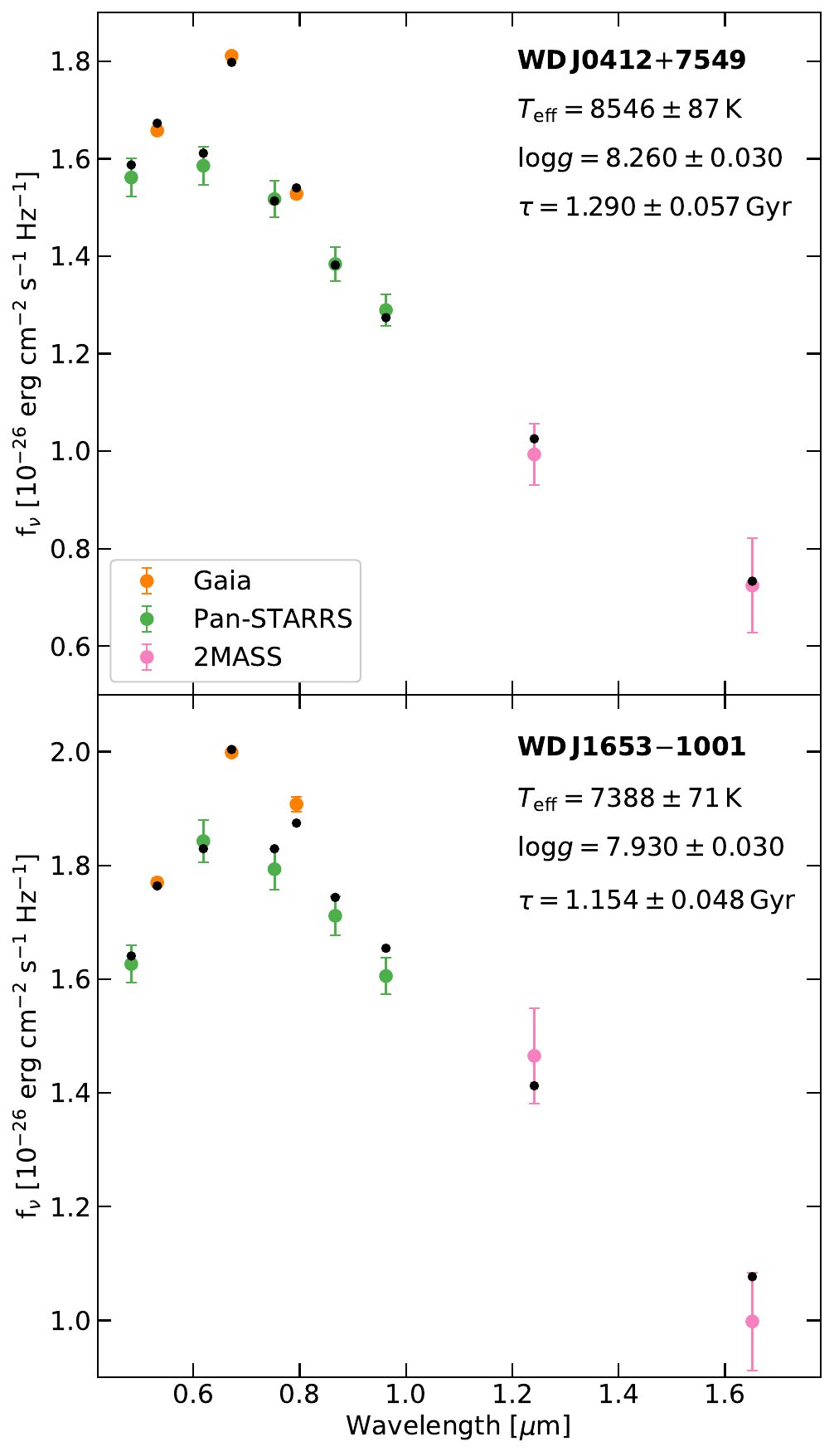}
    \caption{Photometric fits between the observed (coloured) and synthetic (black) photometry for WD\,J0412$+$7549 and WD\,J1653$-$1001. Uncertainties of the Pan-STARRS and 2MASS measurements have an imposed lower limit of 0.025\,mag. Error bars correspond to $1\sigma$ uncertainties.}
\label{fig:WDJ0412_phot_fit}
\end{figure}

Figure~\ref{fig:WDJ0412_phot_fit} shows the spectral energy distribution (SED) created from the photometric fits of WD\,J0412$+$7549 and WD\,J1653$-$1001 with observed and synthetic photometry. Reasonable fits were achieved between the observed and synthetic photometry for both DAe stars. The bandpasses of the \textit{Gaia} filters are broad therefore arbitrary nominal wavelengths were used for the SED. There is no near-IR excess seen in the SED of either DAe star.

We conducted spectroscopic fits using the three DA model atmospheres (like in our photometric fits) on the WHT spectra of WD\,J0412$+$7549 and on the KAST spectra of WD\,J1653$-$1001. The continuum was initially fit on each spectrum with free parameters to extract the Balmer lines. The normalized Balmer lines were then re-fit to give the best-fitting model. We removed H$\alpha$ lines as the emission core made it challenging to accurately fit the spectra. The central regions around the line cores of the other Balmer lines were also removed to achieve the most accurate fit. The WHT observations for WD\,J0412$+$7549 and KAST observations for WD\,J1653$-$1001 have approximately equivalent S/N so the weighted mean of the corresponding sets of \Teff\ and $\log g$ were taken as the spectroscopic atmospheric parameters for each model. Similar to our photometric fits, we added the corresponding uncertainty from the $\approx 2.3$ per cent flux variability of each DAe star to the \Teff\ spectroscopic weighted mean uncertainty. Estimating the uncertainty on the spectroscopic weighted mean of $\log g$ from flux variation is not trivial, so we did not modify those uncertainties.

The 1D convective model provided a worse agreement between photometry and spectroscopy than the more realistic 3D convective model \citep{Tremblay2013} so was discarded. It has been argued that magnetic fields suppress convective energy transfer in white dwarfs \citep{Tremblay2015, Bedard2017, GF2018,Gansicke2020,Cunningham2021} in the regime $\gtrsim 0.01$\,MG at $T_{\rm eff} \approx 8500$\,K, which results in an altered atmospheric temperature structure. Therefore, radiative models are appropriate for stars with $B \gtrsim 0.01$\,MG whereas convective models are best suited for stars with lower magnetic field strengths. We measured the maximum magnetic field strength limit of WD\,J0412$+$7549 and WD\,J1653$-$1001 as $B < 0.05$\,MG therefore either the 1D radiative or 3D convective model could be valid for these stars. 

The 1D radiative spectroscopic fit yielded $\Teff = 8015 \pm 100$\,K and $\log g = 8.305 \pm 0.023$\,dex for WD\,J0412$+$7549 and $\Teff = 6951 \pm 88$\,K and $\log g = 7.622 \pm 0.035$\,dex for WD\,J1653$-$1001. These results are $4\sigma$ and $2\sigma$ away from the photometric \Teff\ and $\log g$ calculated for WD\,J0412$+$7549, and $4\sigma$ and $7\sigma$ away from the photometric \Teff\ and $\log g$ calculated for WD\,J1653$-$1001, respectively.

The 3D convective spectroscopic fit yielded $\Teff = 8578 \pm 106$\,K and $\log g = 8.316 \pm 0.025$\,dex for WD\,J0412$+$7549 and $\Teff = 7613 \pm 95$\,K and $\log g = 7.893 \pm 0.030$\,dex for WD\,J1653$-$1001. These results are $1\sigma$ and $2\sigma$ away from the photometric \Teff\ and $\log g$ calculated for WD\,J0412$+$7549, and $2\sigma$ and $1\sigma$ away from the photometric \Teff\ and $\log g$ calculated for WD\,J1653$-$1001, respectively. 

\begin{figure*}
\centering
\includegraphics[width=0.65\columnwidth]{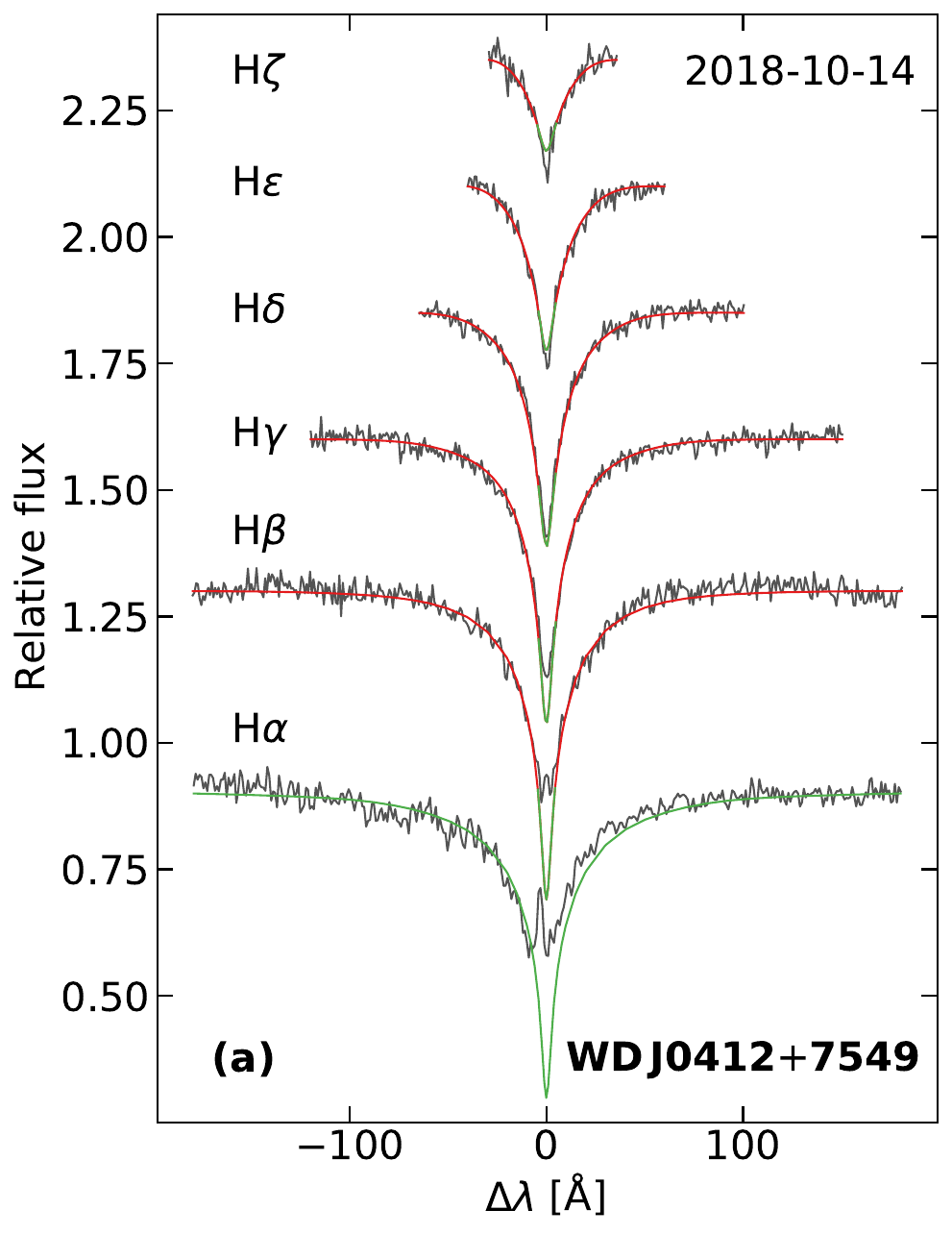}
\includegraphics[width=0.65\columnwidth]{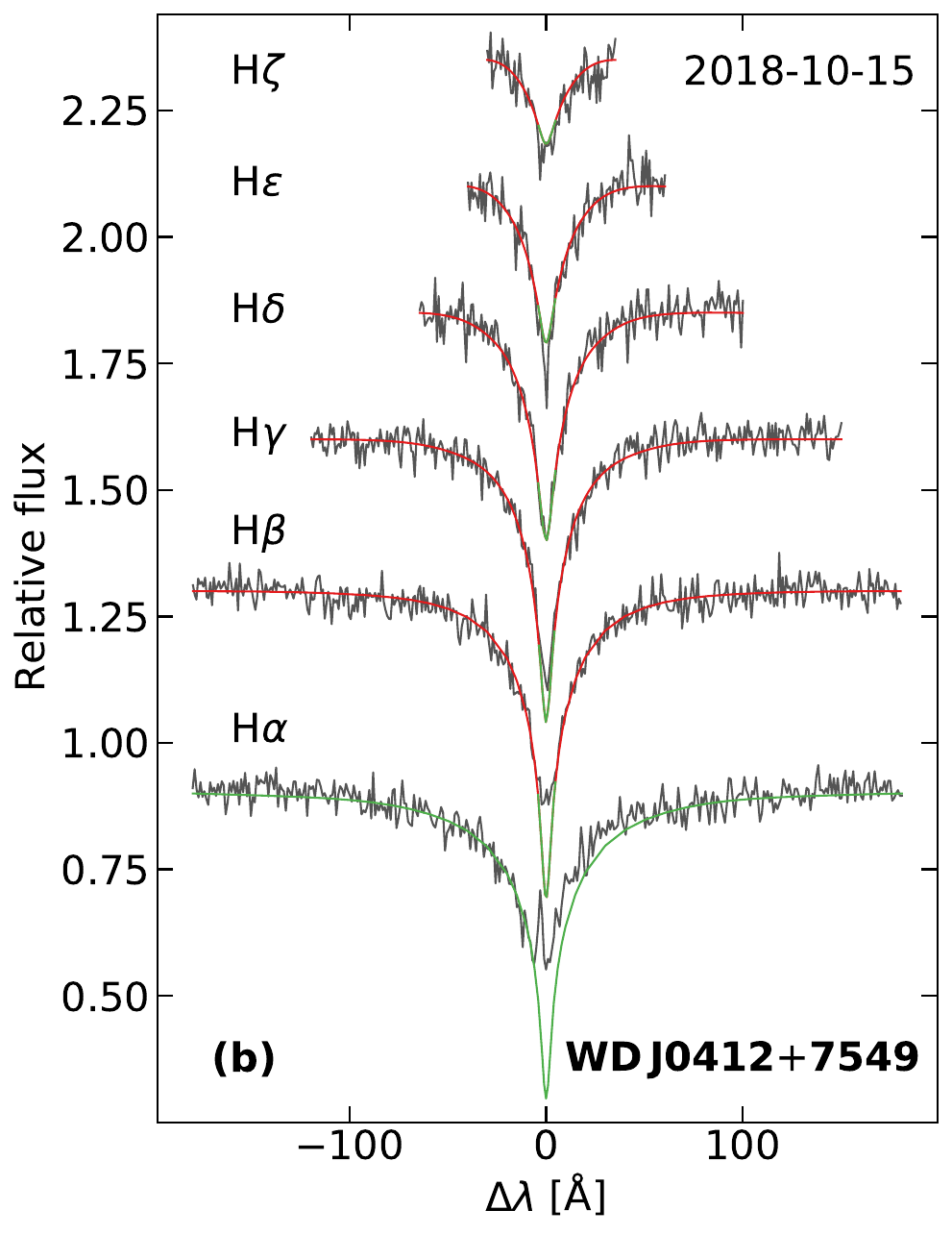}
\includegraphics[width=0.65\columnwidth]{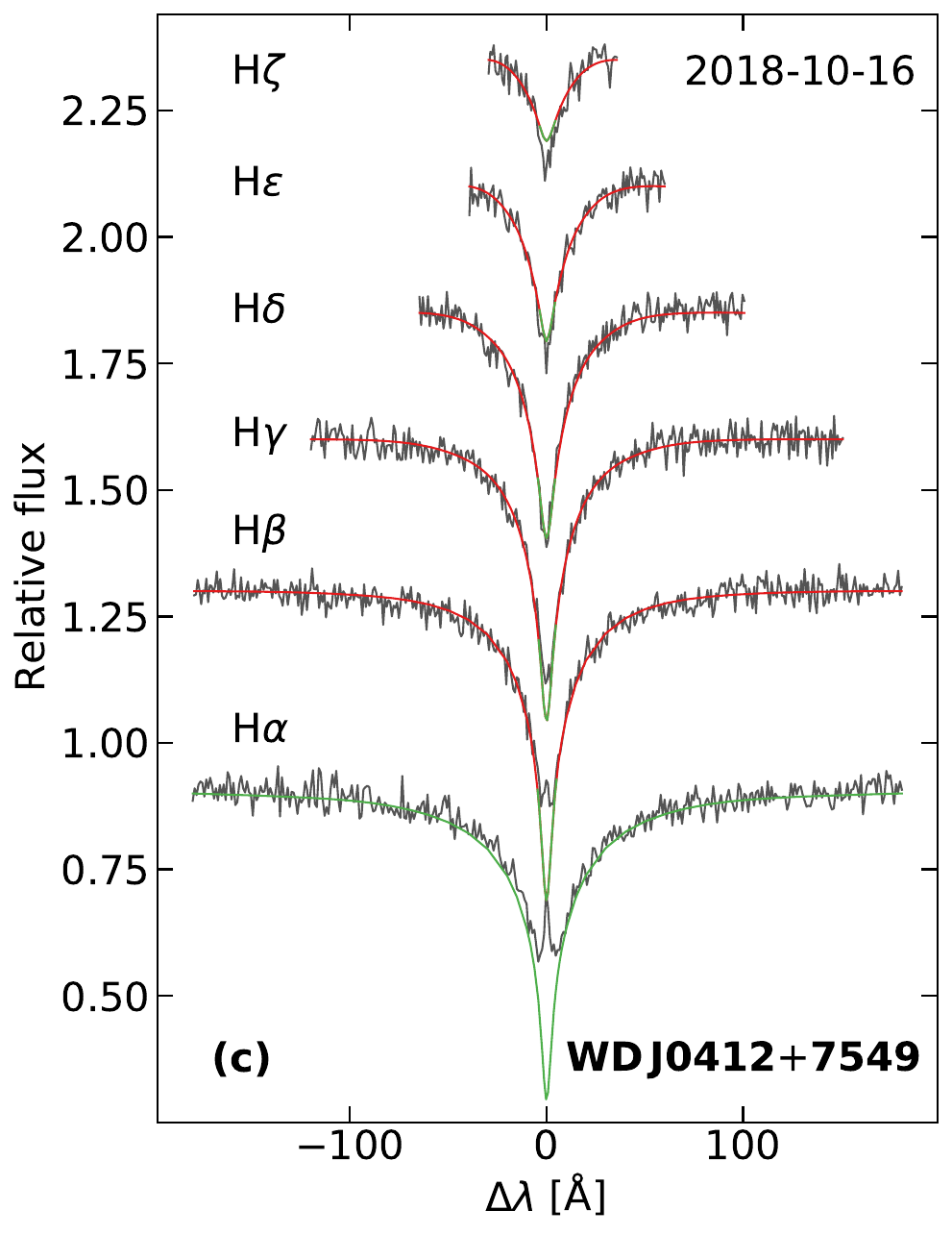}
\includegraphics[width=0.65\columnwidth]{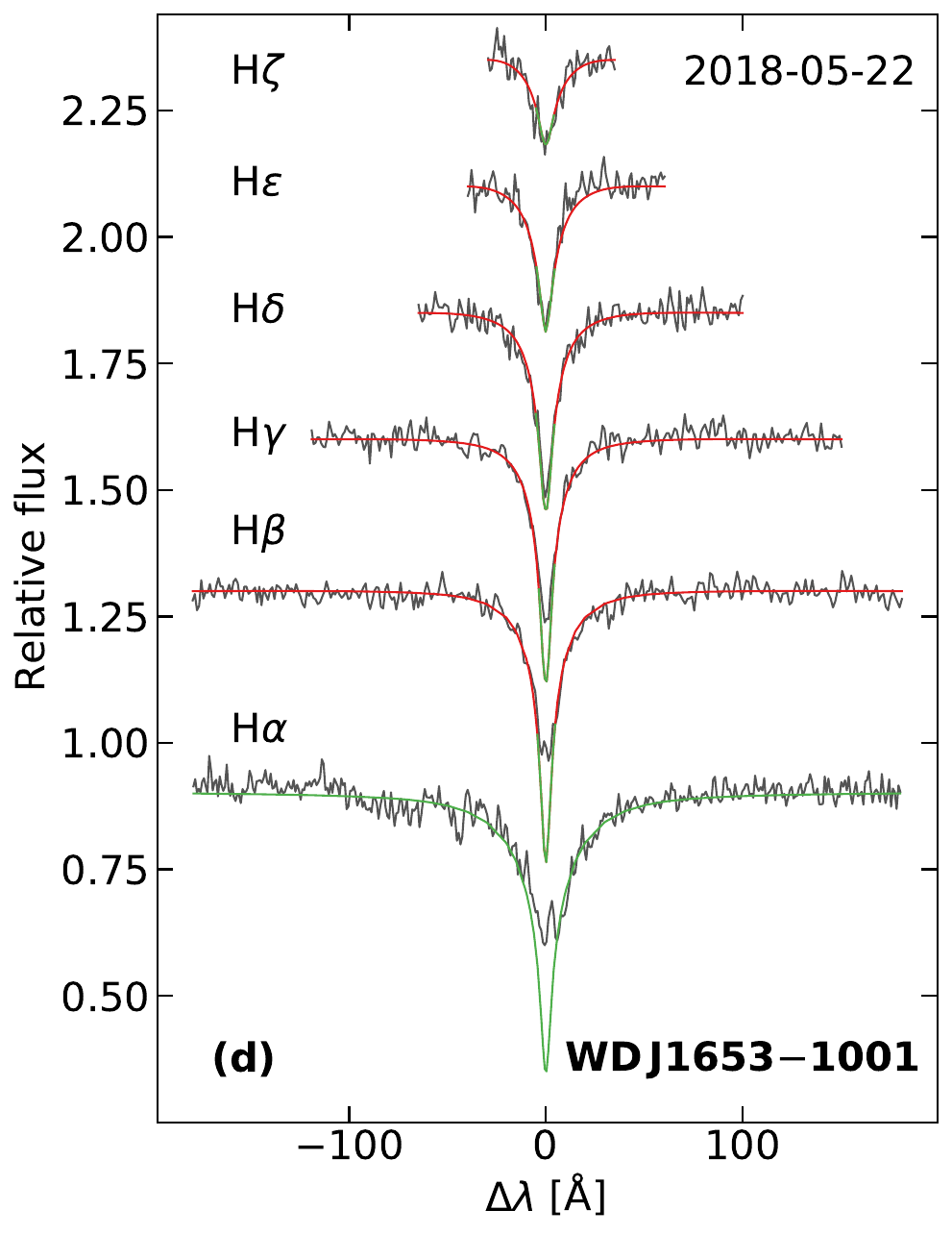}
\includegraphics[width=0.65\columnwidth]{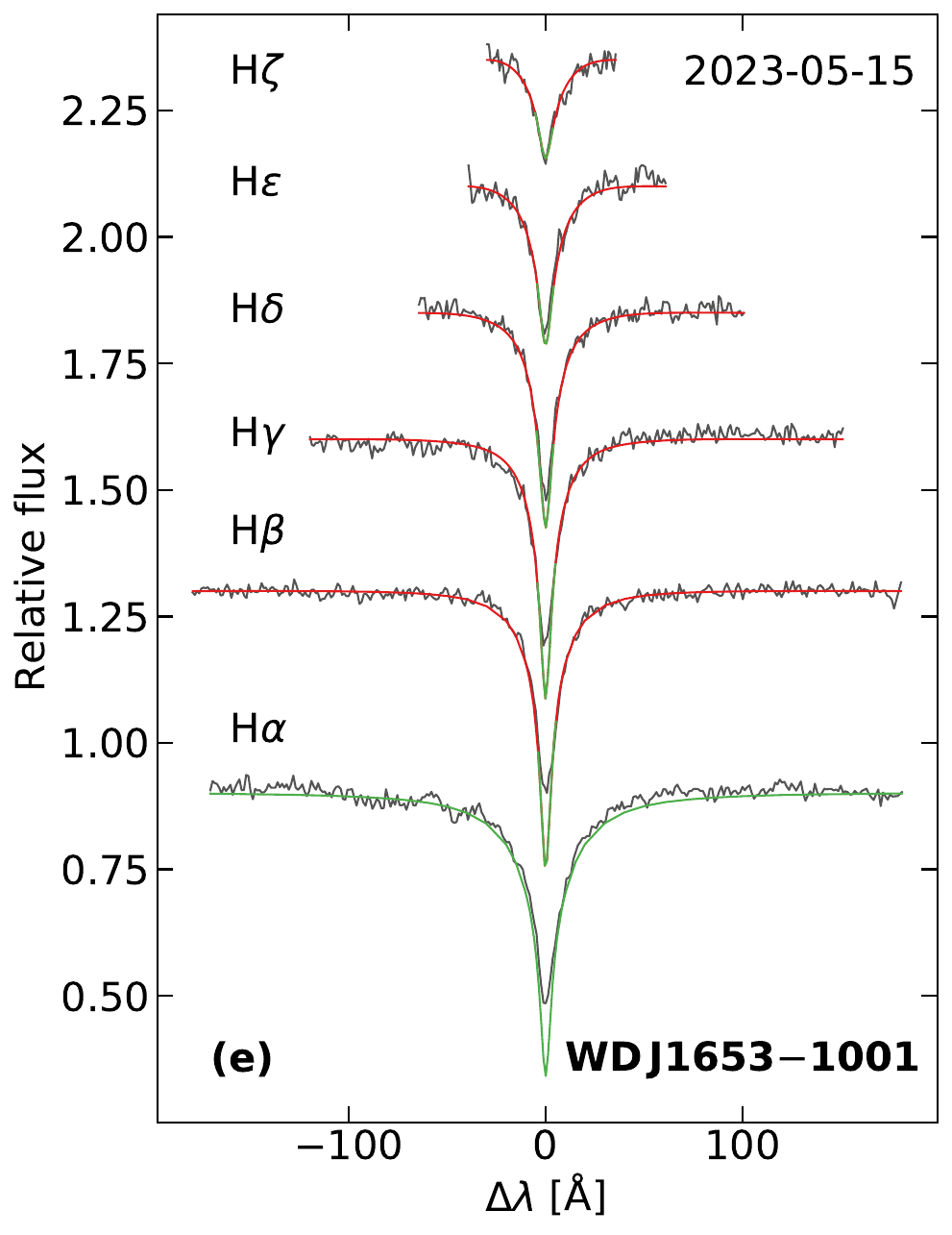}
    \caption{Spectroscopic fits of the normalized observed Balmer line profiles (black) with a 3D convective DA model atmosphere for (a-c) the WHT spectra of WD J0412+7549 and (d-e) the KAST spectra of WD\,J1653$-$1001, where the UT date of the observation is in the top right corner of the plots. The best-fitting model is overlaid in red and the green lines are the regions which were removed from the fit. Lines are offset vertically for clarity.}
\label{fig:spec_fit}
\end{figure*}

\begin{figure*}
\centering
\includegraphics[width=\columnwidth]{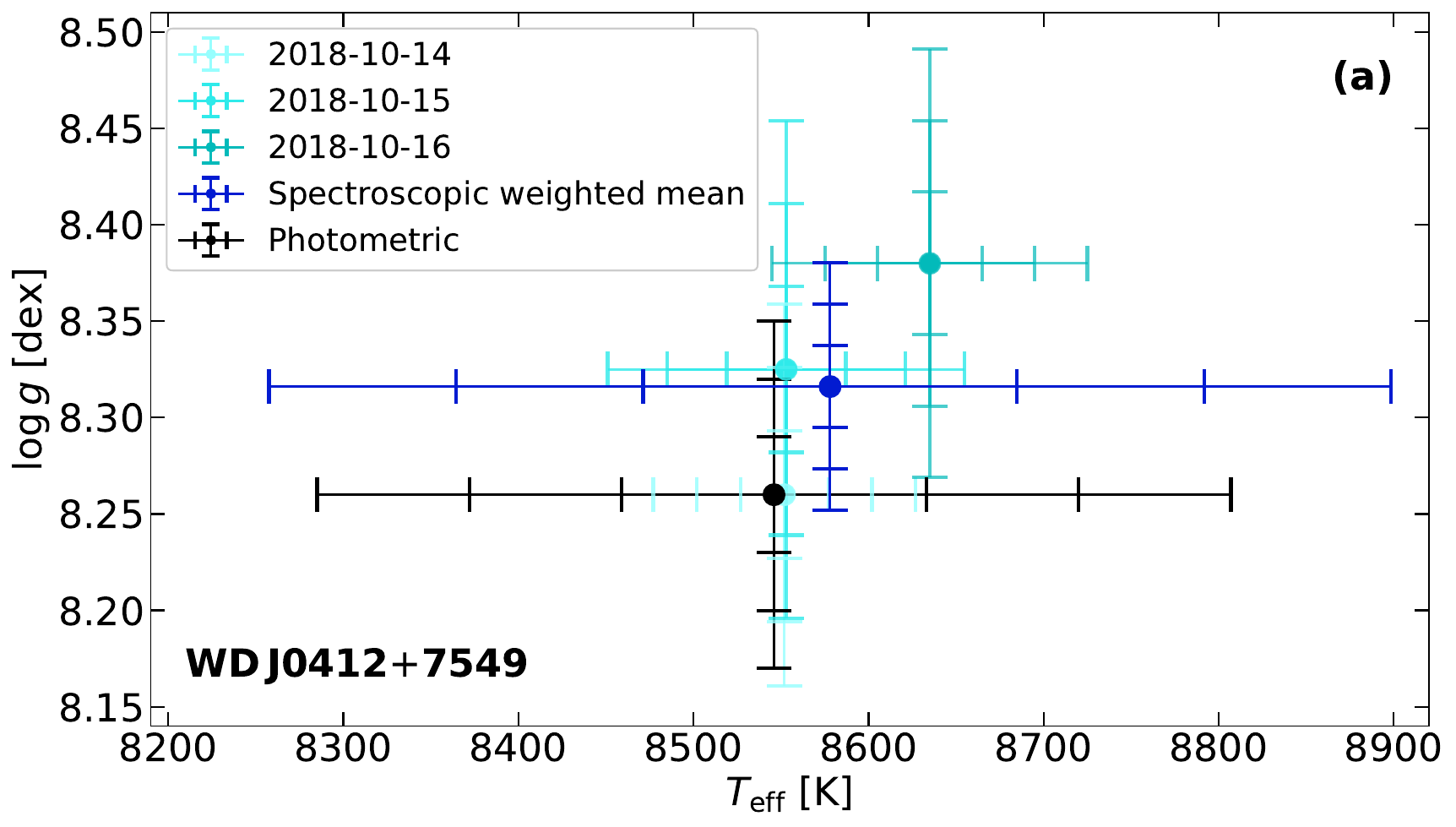}
\includegraphics[width=\columnwidth]{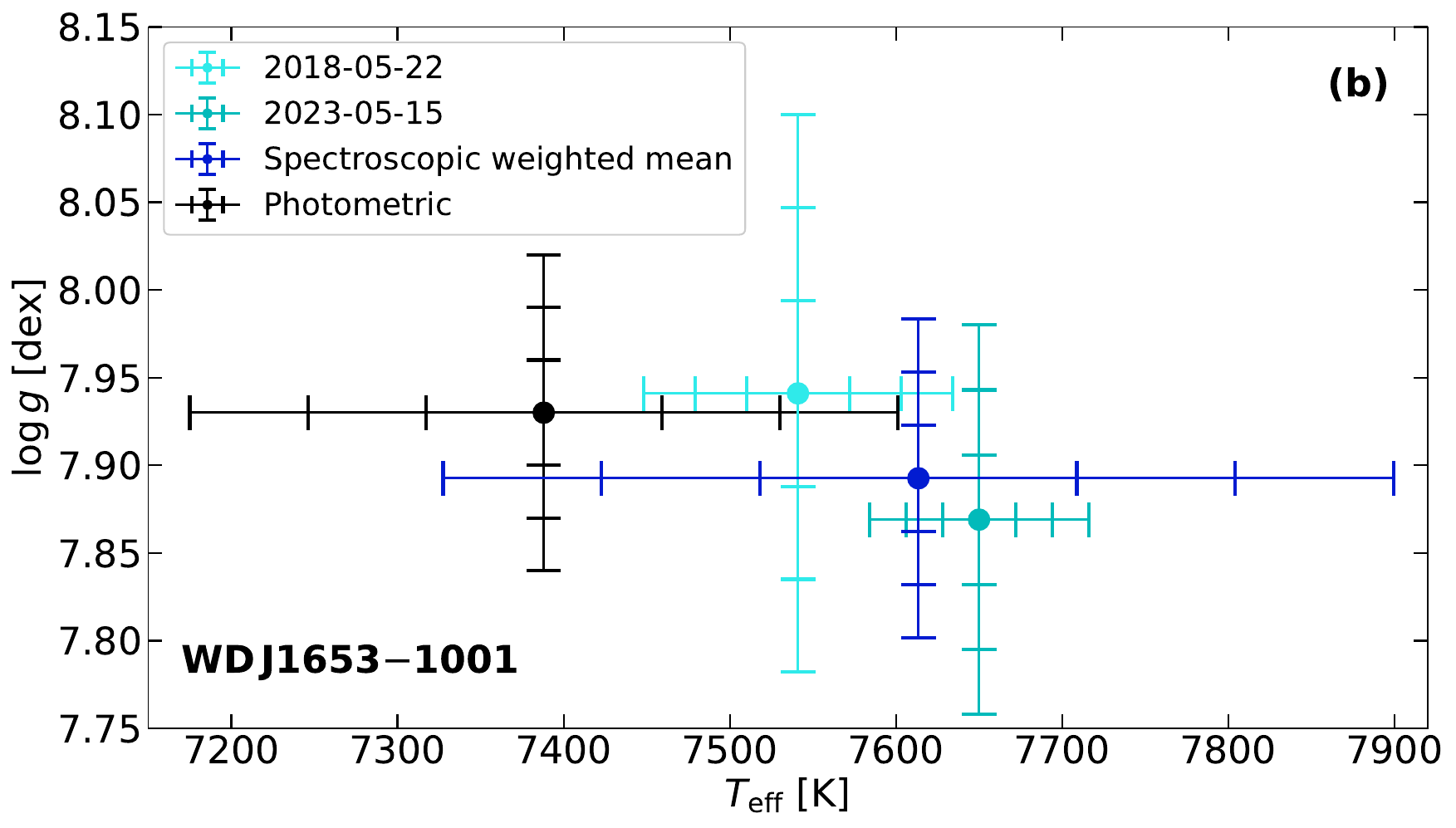}
    \caption{Atmospheric parameters obtained from photometric and 3D spectroscopic fits of (a) WD\,J0412$+$7549 and (b) WD\,J1653$-$1001. 3D spectroscopic fits were performed on each WHT observation of WD\,J0412$+$7549 and KAST observation of WD\,J1653$-$1001, with their corresponding \Teff\ and $\log g$ shown in varying shades of light blue. The 3D spectroscopic weighted mean (dark blue) of the parameters calculated from the individual 3D spectroscopic fits are shown, along with the parameters calculated from the photometric fits (black). Error bars correspond to $1\sigma$, $2\sigma$ and $3\sigma$ uncertainties.}
\label{fig:WDJ0412_phot_spec_fits}
\end{figure*}

The atmospheric parameters obtained using the 3D convective model are in better agreement with the photometric parameters for each star, thus we use 3D convective models for our spectroscopic fits and parameters (Table~\ref{tab:stellar parameters}). The spectroscopic fits using the 3D convective model for the WHT observations of WD\,J0412$+$7549 and KAST observations of WD\,J1653$-$1001 are shown in Figure~\ref{fig:spec_fit}, and Figure~\ref{fig:WDJ0412_phot_spec_fits} shows the corresponding \Teff\ and $\log g$ parameters obtained from these fits in addition to the weighted mean of the 3D spectroscopic fit parameters and the parameters obtained from the photometric fits. 

The photometric parameters for both DAe stars are likely more accurate than the spectroscopic parameters as they are less sensitive to the convection model or emission core removal procedure. Therefore, we use the atmospheric parameters obtained from the photometric fit as our adopted parameters for WD\,J0412$+$7549 and WD\,J1653$-$1001. 

\subsection{Radial velocity of \texorpdfstring{WD\,J0412$+$7549}{WDJ0412+7549}}
\label{sec:radial velocity}

The Balmer emission observed in DAe stars could have intrinsic or extrinsic origins. To test for an extrinsic origin, we searched for evidence of a companion using the radial velocity ($v_{\rm rad}$) of the H$\alpha$ emission line core from multiple spectroscopic exposures. We measured the wavelength shift of the H$\alpha$ emission line central components in WD\,J0412$+$7549 compared to rest wavelengths in all Keck, INT and Gemini spectroscopic data. The spectra observed using Keck are of the highest spectral resolution compared to the other observations so are the most reliable to calculate the radial velocity of WD\,J0412$+$7549. A reliable radial velocity variation of WD\,J1653$-$1001 cannot be calculated with the available KAST spectra. 

Our procedure consisted of creating a Gaussian of a fixed-width and convolving it by the spectral resolution of the instrument at the H$\alpha$ line. We then fitted the Gaussian to the emission line 10\,000 times while moving it horizontally within the noise of the core emission wings. The peak of the fitted Gaussian which corresponded to the minimum convolution between the Gaussian and emission line was the best-estimate of the central wavelength of the emission line. The Keck spectroscopic data were reduced with MAKEE and output with vacuum wavelengths corrected to the heliocentric reference frame, but as no further corrections were made we imposed a lower limit of 5\,\kms\ for radial velocity uncertainties. For consistency, we applied the same lower limit for all radial velocity uncertainties. 
 
The line velocity obtained by the first Keck observation taken at 08:22:08.25\,UT was $21.2 \pm 6.5$\,\kms. The second Keck observation was taken at 12:09:41:97\,UT, which is at a 0.438 phase shift from the end of the first exposure. The H$\alpha$ core emission line variability between exposures is significant, with the second Keck exposure evidencing a much shallower emission core (Figure~\ref{fig:Keck_Ha_Hb}). The H$\alpha$ emission line in the first Keck exposure has an approximate Gaussian structure thus our fitting procedure to measure the line velocities was successful. Accurately measuring the H$\alpha$ line velocity in the second Keck exposure proved challenging because the emission drastically decreases and the S/N ratio does not allow us to confidently say if there is sub-structure within the emission lines. Since the earlier Keck exposure clearly showed the emission line cores to have a Gaussian structure, we assumed the core in the second exposure does too. We therefore performed the same fitting procedure on the second Keck exposure to yield the H$\alpha$ emission line velocity of $14.3 \pm 7.2$\,\kms. 

The gravitational redshift of WD\,J0412$+$7549 was calculated to be $45.1 \pm 1.5$\,\kms. Correcting the emission line velocities for the gravitational redshift yielded $v_{\rm rad}$ of $-23.9 \pm 6.6$\,\kms\ and $-30.7 \pm 7.4$\,\kms\ for the first and second Keck exposures, respectively. The direction of movement of WD\,J0412$+$7549 is unknown so we only consider the radial velocity variation. Therefore, the radial velocity variation of WD\,J0412$+$7549 is $6.9 \pm 9.9$\,\kms\ for a phase difference of 0.438, which is consistent with zero. Using the same fitting procedure on all Gemini spectra, we get an upper limit on the radial velocity variation of $\lesssim 40$\,\kms.

\subsection{Core crystallization}
\label{sec:crystallization}

The onset of crystallization has been suggested as a potential scenario for the close clustering of DA(H)e stars on the \textit{Gaia} HRD (see Section~\ref{sec:physical characteristics}) and for the production of their magnetic fields through a crystallization-driven convective dynamo \citep{Isern2017, Schreiber2021a, Schreiber2021b}.

We investigated whether the two DAe stars have started crystallising and, if so, what percentage of their interior is crystallized. To this end, we computed new model evolutionary sequences using the STELUM code \citep{Bedard2022}. Our white dwarf models consist of a carbon/oxygen/$^{22}$Ne core, surrounded by a standard helium mantle ($\mathrm{M}_{\mathrm{He}}/\mathrm{M}_{\mathrm{WD}} = 10^{-2}$) and a "thick" outermost hydrogen layer ($\mathrm{M}_{\mathrm{H}}/\mathrm{M}_{\mathrm{WD}} = 10^{-4}$). The core composition is initially uniform, with a standard $^{22}$Ne mass fraction of 1.4 per cent. As for the oxygen mass fraction, the appropriate value is still quite uncertain, which is unfortunate given that this parameter significantly affects the crystallization process (i.e. a more oxygen-rich white dwarf starts to crystallize earlier; \citealt{Fontaine2001,Bauer2020}). On one hand, current pre-white dwarf evolutionary models typically predict central oxygen abundances between 60 and 70 per cent \citep{Renedo2010,Salaris2022,Bauer2023}. However, these predictions are notoriously uncertain due to our poor knowledge of nuclear reaction rates and convective boundary mixing in the core helium burning phase \citep{Straniero2003,Salaris2010,DeGeronimo2019,Giammichele2022}. On the other hand, recent asteroseismological analyses of pulsating white dwarfs hint that the central oxygen abundance may be closer to (and perhaps even higher than) 80 per cent \citep{Giammichele2018,Giammichele2022}. To account for this uncertainty, we computed two sets of sequences assuming core oxygen mass fractions of 60 and 80 per cent, respectively.

Our new sequences also incorporate several notable improvements at the level of the input physics with respect to the sequences published in \citet{Bedard2020}. We included the energy released by $^{22}$Ne diffusion in the liquid phase and by carbon/oxygen phase separation upon crystallization, as outlined in \citet{Bedard2022} and with minor updates discussed in \citet{Venner2023}. We also made use of the improved carbon/oxygen phase diagram of \citet{Blouin2021}, which is a critical ingredient to accurately model the crystallization process. Finally, we employed the envelope conductive opacities of \citet{Blouin2020_opac}, which result in a faster cooling and thus a slightly earlier onset of crystallization with respect to the previous conductive opacities of \citet{Cassisi2007}.

\begin{figure*}
\centering
\includegraphics[width=\columnwidth]{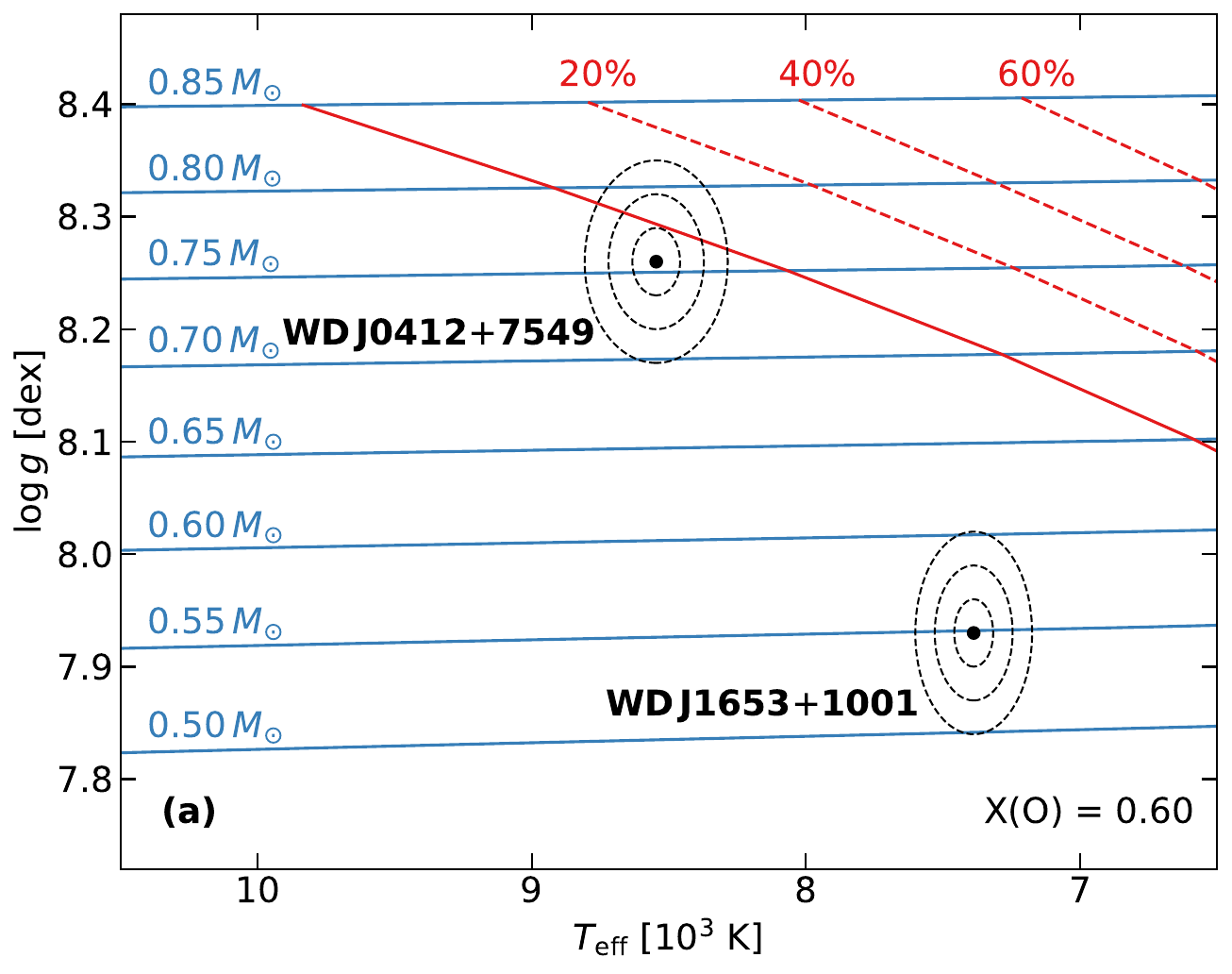}
\includegraphics[width=\columnwidth]{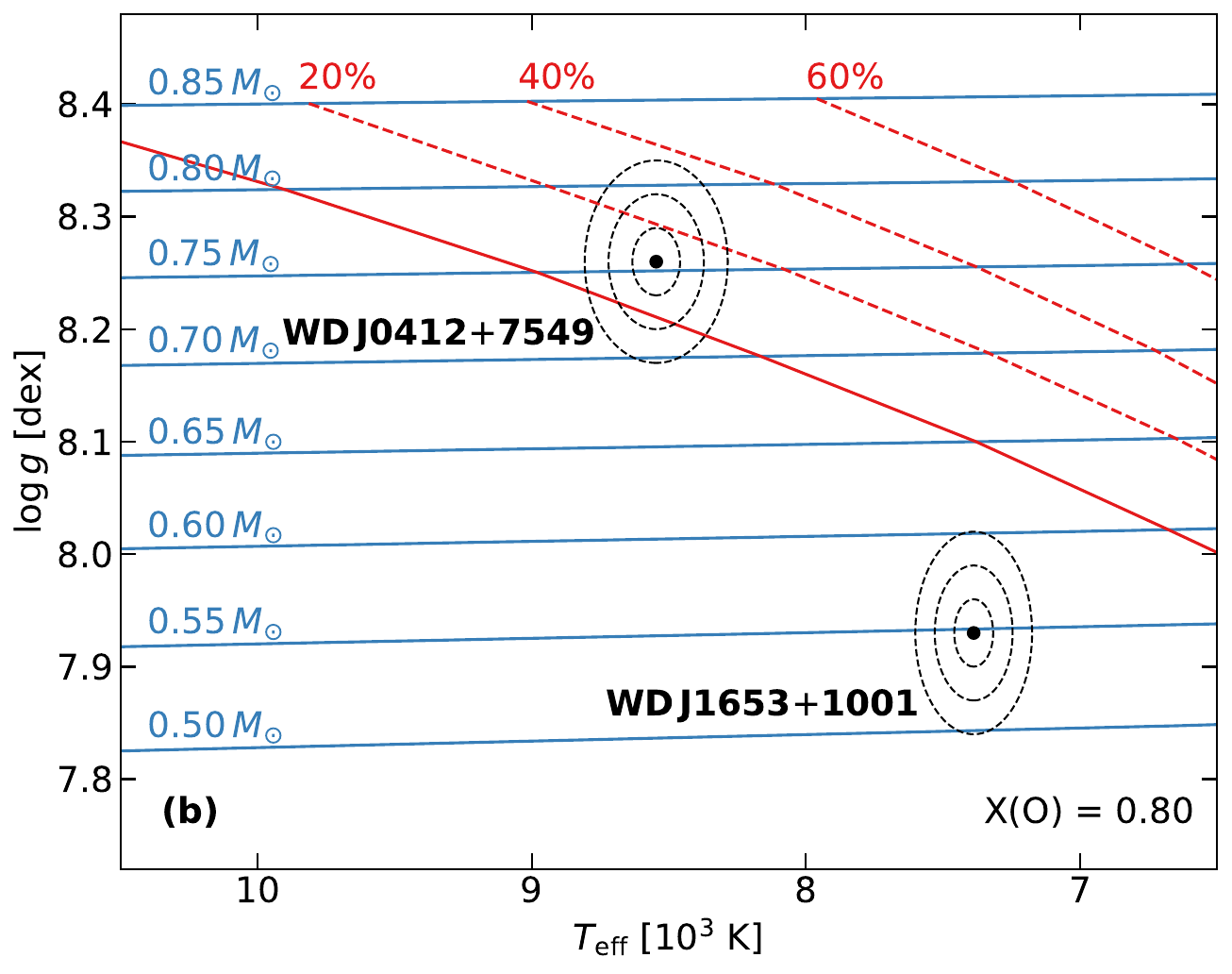}
    \caption{Evolutionary models showing the crystallized fraction (by mass) for WD\,J0412$+$7549 and WD\,J1653$-$1001 assuming core oxygen mass fractions of (a) 60 per cent and (b) 80 per cent. The photometric parameters of the two DAe stars are represented by black dots, with the dotted ellipses showing the $1\sigma$, $2\sigma$ and $3\sigma$ uncertainties. Shown are the model sequences for white dwarf masses $0.50 - 0.85$\,\Msun\ (blue tracks), the onset of crystallization (solid red line) and the evolutionary stages where the star is 20 per cent, 40 per cent or 60 per cent crystallized (dotted red lines).}
\label{fig:DAe_crystallization}
\end{figure*}

Figure~\ref{fig:DAe_crystallization} shows the predicted crystallized fraction (by mass) for WD\,J0412$+$7549 and WD\,J1653$-$1001 according to our model evolutionary sequences. Assuming a core oxygen abundance of 60 per cent, crystallization has not started for WD\,J0412$+$7549 below 1$\sigma$. At 3$\sigma$, the crystallized fraction could reach $\approx 15$ per cent. However, if we assume a core oxygen abundance of 80 per cent for WD\,J0412$+$7549, then the predicted crystallized fraction is $\approx 15$ per cent. At 1$\sigma$ the crystallized fraction could increase to 20 per cent, and further increase to almost 40 per cent at 3$\sigma$. Crystallisation has not started for WD\,J1653$-$1001 up to 3$\sigma$ when assuming either core oxygen abundances.

\section{Discussion}
\label{sec:Discussion}

\subsection{Physical characteristics of DA(H)e stars}
\label{sec:physical characteristics}

WD\,J0412$+$7549 and WD\,J1653$-$1001 lack observable Zeeman-split Balmer lines in their emission cores upon inspection of their spectra (see Secion~\ref{sec:Spectroscopic variability}), hence have no detectable magnetic field and are classified as DAe. This is an important difference with the larger DAHe class as these stars have Zeeman-split Balmer lines with measured field strengths in the range $\simeq 5 - 147$\,MG \citep{Greenstein1985, Gansicke2020, Reding2020, Walters2021, Manser2023, Reding2023}, which are $2-3$ orders of magnitude higher than the upper limit determined for DAe white dwarfs. There are known magnetic white dwarfs with field strengths between $0.05 - 5$\,MG, e.g. in the 40\,pc sample \citep{BagnuloLandstreet2021,Bagnulo2022,OBrien2023} so possible reasons for the apparent absence of DA(H)e stars in this field range are: poor spectroscopic phase-coverage of magnetic DA stars resulting in the misclassification of DA(H)e stars as DAH or DA due to phase-dependent emission \citep{Manser2023}; the emission strength may correlate with magnetic field strength, making the identification of Zeeman-split Balmer emission lines more difficult at low magnetic field strengths and in spectra which are not of high S/N or resolution \citep{Bagnulo2018, Ferrario2020}. Dedicated and high-resolution spectroscopic follow-up of all white dwarfs in the relevant portion of the \textit{Gaia} HRD is required to understand whether this gap of field strengths in DA(H)e white dwarfs is real or a selection effect.

\begin{figure}
\centering
\includegraphics[width=\columnwidth]{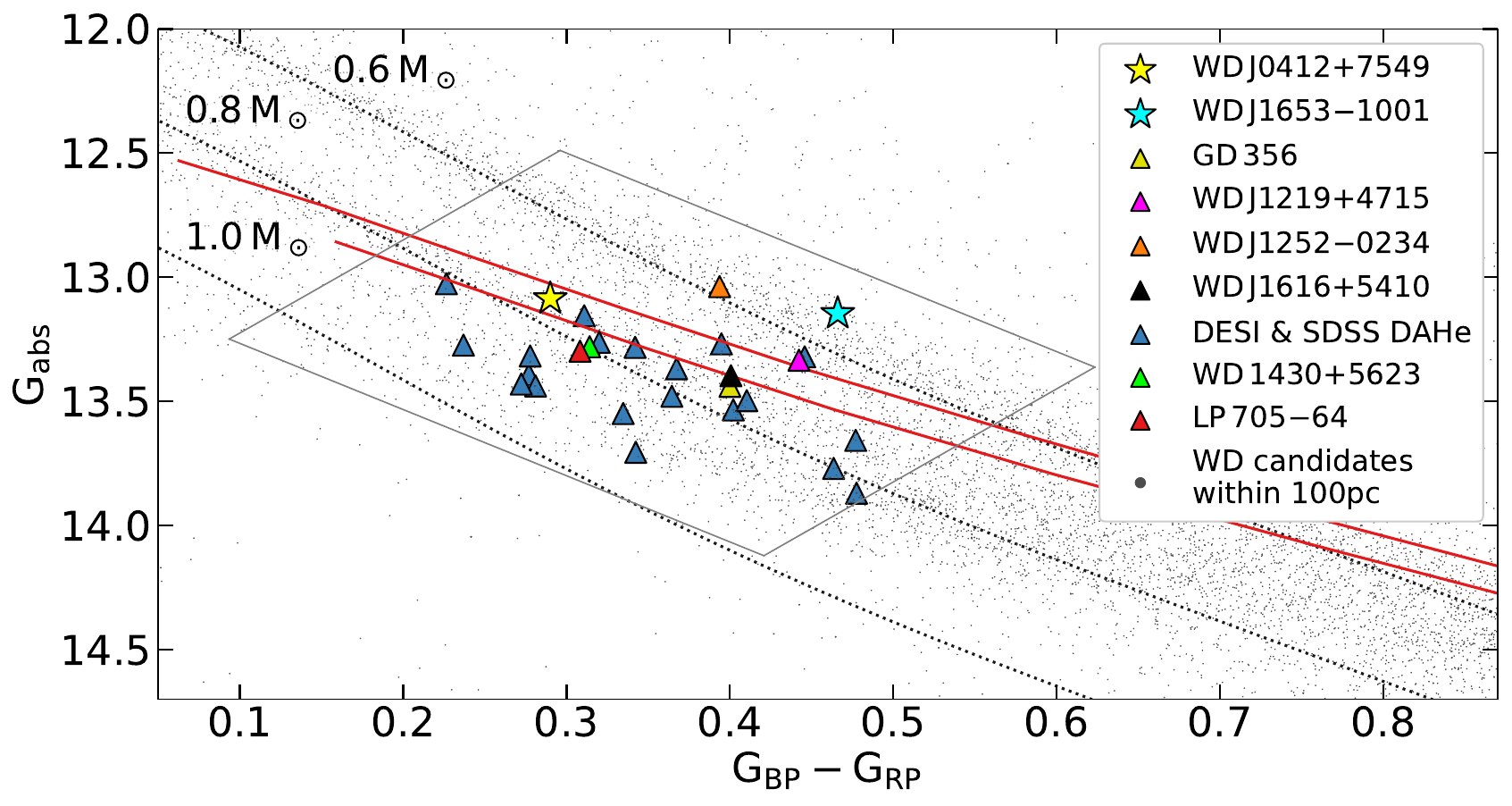}
    \caption{\textit{Gaia} HRD zoomed-in on the white dwarf cooling sequence where DA(H)e reside, which is based off fig.~\,8 in \citet{Manser2023}. The two DAe stars WD\,J0412$+$7549 and WD\,J1653$-$1001 (stars) are shown, in addition to the DAHe white dwarfs from the literature (triangles), compared to white dwarf candidates within 100\,pc of the sun \citep[grey dots;][]{GF2019}. The grey box shows the region defined in \citet{Manser2023} which bounds the DA(H)e stars. The black dotted lines show the cooling tracks of 0.6\Msun, 0.8\Msun\ and 1.0\Msun\ DA white dwarfs calculated with new model evolutionary sequences (discussed in Section~\ref{sec:crystallization}) and a core oxygen mass fraction of 60 per cent. The red solid lines show the predicted onset of crystallization assuming core oxygen mass fractions of 60 per cent (lower) and 80 per cent (upper).}
\label{fig:HRD}
\end{figure}

However, DAe stars have undeniable similarities to the larger DAHe class. None of the known DA(H)e stars exhibit metal absorption features in their spectra~--~even high-resolution spectra of WD\,J0412$+$7549 does not show evidence of metal polution, as predicted in \citet{Walters2021}~--~which suggests they are not actively accreting nor have recently accreted planetary material \citep{Koester1997, Jura2003,Gansicke2019}. Also, all DA(H)e stars closely cluster in one region of the \textit{Gaia} HRD in $G_{\mathrm{BP}} - G_{\mathrm{RP}}$ versus $G_{\mathrm{abs}}$ compared to the parameter space occupied by white dwarf candidates within 100\,pc of the sun \citep[Figure~\ref{fig:HRD};][]{GF2019}. \citet{Manser2023} explored the close clustering of DA(H)e stars and found that within the cluster region $\approx 3$ per cent of white dwarfs are classified as DA(H)e and $\approx 10 - 30$ per cent of DAH stars exhibit Balmer line emission. The close clustering could suggest that Balmer line emission is a consequence of a short-lived evolutionary stage of DAH or DA white dwarfs which have cooling ages of $>1$\,Gyr \citep{Manser2023}. The cause of the trigger of Balmer line emission in this niche evolutionary stage is still unknown.

There is a homogeneity in atmospheric parameters of the DA(H)e stars as they have $7400\,\mathrm{K} \lesssim \Teff \lesssim 8500$\,K and white dwarf masses $0.5\Msun \lesssim \mathrm{M}_{\mathrm{WD}} \lesssim 0.8$\,\Msun. The spin period of WD\,J0412$+$7549 is consistent with those of DAHe stars which fall in the range $\simeq 0.08 - 36$\,h. We derived a tentative spin period of $\simeq 80.5$\,h for WD\,J1653$-$1001 (Section~\ref{sec:Photometric variability of WDJ1653}) which is slower than the spin periods of other DA(H)e stars. We confirm spectral variability for WD\,J1653$-$1001, as we find evidence of Balmer line cores filled with emission in all observed phases (Figures~\ref{fig:KAST_Ha_Hb_Hg_Hd_Hz}~and~\ref{fig:spec_fit}), but additional time-domain observations and analysis is required to unambiguously determine the variability nature of WD\,J1653$-$1001. Out of the 28 DA(H)e stars, 57 per cent are photometrically variable and 29 per cent are spectroscopically variable. Note that these variability fractions are likely lower limits since not all DA(H)e have time-domain data. 

It is clear that for WD\,J0412$+$7549 the photometric flux maximum corresponds to the minimum strength of emission lines, i.e. the photometric and spectroscopic variability are in anti-phase (Section~\ref{sec:Spectroscopic variability}) - which is the relationship predicted in \citet{Walters2021} for this star. The same phase relationship was found in the DAHe stars SDSS\,J125230.93$-$023417.72 \citep{Reding2020}, GD\,356 \citep{Walters2021}, WD\,J1616$+$5410 \citep{Manser2023}, suspected in SDSS\,J121929.45+471522.8 \citep{Gansicke2020} and visually identified in LP\,705$-$64 and WD\,J143019.29$-$562358.33 \citep{Reding2023}. Thus, the seven DA(H)e stars that have adequate phase coverage in spectroscopic observations and exhibit spectroscopic variability have a confirmed or suspected anti-phase relationship with photometric variability. An anti-phase relationship is indicative of a dark spot/region beneath an optically thin chromosphere, whereas we would expect an in-phase relationship if there is a closely orbiting companion. 

WD\,J0412$+$7549 has radial velocity variation of $\lesssim 40$\,\kms\ and both DAe stars have no evidence of near-IR excess, suggesting that neither star has a stellar companion (see Section~\ref{sec:Unseen companions}). This is consistent with the DA(H)e class in general, where none of the class members have confirmed stellar or planet mass companions.

\subsection{What is causing the \texorpdfstring{H$\alpha$}{Ha} and \texorpdfstring{H$\beta$}{Hb} emission lines?}
\label{sec:emission reason}

Both non-magnetic and magnetic mechanisms have been put forward as possible reasons for Balmer emission lines in cool white dwarfs ($\Teff \leq 8500$\,K). In the following subsections, we will discuss the feasibility of each of these mechanisms being present in the DAe stars WD\,J0412$+$7549 and WD\,J1653$-$1001.

\subsubsection{Unseen companions}
\label{sec:Unseen companions}

We now explore the possibility of a stellar or substellar companion, such as a brown dwarf or planet, orbiting the DAe stars and causing the Balmer emission lines. It is feasible that a white dwarf can have a planetary mass companion as it is well established that planets can survive the stellar evolution of their host star into a white dwarf \citep{Villaver2007, Mustill2012, Rao2018, Ronco2020}. Recent observations have suggested planet candidates orbiting white dwarfs \citep{Thorsett1993, Sigurdsson2003, Luhman2011, Gansicke2019, Vanderburg2020, Blackman2021}, but no substellar companions have been confirmed around DA(H)e stars at this time. 

We are able to place upper mass limits on potential companions of WD\,J0412$+$7549 and WD\,J1653$-$1001 by comparing infrared photometry of these stars from 2MASS, JHK and \textit{WISE} reported in the CatWISE2020 catalogue \citep{Marocco2021} to brown dwarf flux models \citep{Phillips2020}. Despite neglecting background contaminated \textit{WISE} photometry from all photometic fits performed in this work, we include it here as the peak wavelengths of late-spectral-type objects fall in the far-IR thus \textit{WISE} measurements are optimal for placing limits on potential brown dwarf companions. The $W1$ band places the strongest constraints on companion spectral type due to a flux dip in brown dwarf models in $W2$. Using the $\mathrm{M}_{\mathrm{WD}}$ we derived from photometry (Table~\ref{tab:stellar parameters}) and the initial-to-final-mass relation (IFMR) from \citet{Cummings2018}, we calculated the progenitor mass of WD\,J0412$+$7549 to be $3.1 \pm 0.2$\,\Msun\ and the total age of the system to be $1.54 \pm 0.13$\,Gyr. The photometry in the $W1$ filter is $15.226 \pm 0.023$\,mag which, with the total age of the system, places a mass constraint on a substellar companion no earlier than a T-type brown dwarf ($< 60$\,\MJup). Doing the same analysis for WD\,J1653$-$1001 reveals a progenitor mass of $0.8 - 0.9$\,\Msun\ and the total age of the system to be $> 10$\,Gyr. The $W1$ measurement for this system is $14.347 \pm 0.015$\,mag but there is no evidence that the white dwarf and close-proximity background main-sequence star have been resolved. Using the $W1$ measurement and total age of the system, we can place a mass constraint of $< 80$\,\MJup\ on a companion, which rules out a stellar companion and anything earlier than a brown dwarf. Even though the $W1$ measurement is contaminated, we reach a similar result when using 2MASS photometry from the $JHK$ filters.

If the Balmer line emission in these DAe stars originates from a companion, then we expect the orbital period to be the same as the photometric variability period. Kepler's laws can be used to predict the radial velocity of the emission feature if it was emitted by a companion in WD\,J0412$+$7549. We do not entertain an inclination of 0$^{\circ}$ (face-on orbit) as we would not observe photometric variations in this case. Assuming an inclination of 90$^{\circ}$ (edge-on orbit), the companion would require a radial velocity variation with a lower limit of $\gtrsim 400$\,\kms\, compared to an observed radial velocity variation upper limit of $\approx 10$\,\kms\ for WD\,J0412$+$7549. Even an inclination of 45$^{\circ}$ would require the companion to have a radial velocity of $\approx 280$\,\kms\, which is still infeasibly large. Therefore, we can largely rule out that the emission is from a companion and instead it must originate from the stellar surface.

If DAe stars have a companion but the Balmer emission lines are assumed to originate from the white dwarf surface and the orbital period corresponds to the photometric period, then the radial velocity of the star will be dependent on the companion mass. During the companion's orbit, the force exerted onto the star would cause the Balmer line emission cores and wings to radially shift with a consistent amplitude. With the current observations of WD\,J0412$+$7549, it is only possible to measure the radial velocity of the Balmer line emission cores but this is sufficient for our analysis. We can constrain a companion upper mass limit of $115$\,M$_\oplus$ for a favourable edge-on inclination.

\subsubsection{Magnetic mechanisms}
\label{sec:magnetic mechanisms}

It is possible that the same magnetic mechanism which causes the detectable magnetic field in DAHe stars is present in DAe stars and causes all objects in the DA(H)e class to have Balmer emission lines, yet the magnetic field in DAe stars is not strong enough to cause these lines to be Zeeman-split into triplets. Magnetic field lines could emerge for a certain \Teff\ and $\mathrm{M}_{\mathrm{WD}}$, resulting in an intrinsically-activated chromosphere for a limited amount of time, before the white dwarf continues to evolve along the cooling track as a DA or DAH. The undetectable magnetic field in DAe stars could therefore be due to them being at slightly different stages of the DA(H)e evolutionary phase which could cause the magnetic field to be smaller or buried below the photosphere as the field lines simply have not had time to emerge from the surface yet.

The origin of magnetic fields in isolated degenerate stars is not well understood, although several theories have been developed to explain their presence, such as: fields remnant from the progenitor pre-main sequence (fossil fields) or main sequence stars that got trapped and retained in the non-convective core regions of the star and released upon the evolution into a white dwarf \citep{Landstreet1967, Angel1981, Braithwaite2004, Tout2004, Wickramasinghe2005}; a dynamo acting in the common envelope phase \citep{Tout2004, Briggs2018, Belloni2020} or during a merger \citep{Garcia2012}; or a convective dynamo driven by white dwarf core crystallization \citep{vanHorn1968, Isern2017, Schreiber2021a, Schreiber2021b, Ginzburg2022}. 

The unipolar inductor model has been explored in the literature \citep{Goldreich1969, Li1998, Wickramasinghe2010} as a potential mechanism for magnetism and the emission feature in DAHe stars \citep{Gansicke2020, Reding2020, Walters2021}. This model involves the induction of an electric current from the close-in orbit of a rocky planet through the host star's magnetosphere, which consequently heats up the host's atmosphere at the magnetic poles causing emission. An in-phase relationship between photometric and spectroscopic variability is expected from the unipolar model. However, DA(H)e stars have been found to have an anti-phase relationship (Section~\ref{sec:Spectroscopic variability}), suggesting they host a photospheric dark spot below an optically thin emission region, i.e. with a temperature inversion above the photosphere, which is inconsistent with the unipolar model \citep{Walters2021}.

Observational evidence and the similar physical characteristics of DA(H)e stars suggest that the Balmer emission lines are caused by a mechanism internal to the white dwarf, such as magnetic emergence or an intrinsically-activated chromosphere. An active and hot chromosphere (above the photosphere) in polar magnetic regions would cause the change in intensity of Balmer emission lines which we observe as spectroscopic variability over the spin period, in addition to flux variations from underlying cooler/warmer photospheric regions which we observe as photometric variability as the white dwarf rotates. Note that photometric variability could also be caused by other magnetic effects (opacities, polarization) and more observations (e.g. multi-wavelengths) are needed to conclude whether the photometric temperature is variable over the surface \citep{Fuller2023}.

A local surface dynamo (or chromospheric activity itself) to explain magnetic field generation is ruled out as the amount of energy stored in the convection zone and upper layers is unable to explain magnetic fields larger than about 1\,kG \citep{Fontaine1973,Tremblay2015}. 

A crystallization-induced global dynamo has been theorized to cause the production and emergence of magnetic fields in isolated white dwarfs thus could be the reason for magnetism in DA(H)e stars \citep{Isern2017, Schreiber2021a, Schreiber2021b}. The Balmer line emission could then result from the emergence at the stellar surface of these newly generated magnetic field lines. But, more recently, doubts have been raised on the efficiency of this mechanism due to the small convective velocities\footnote{This refers to slow compositionally-driven internal convection and it is unrelated to the surface dynamo discussed earlier.} and kinetic energy flux reservoir \citep{Fuentes2023}. The onset of core crystallization depends on white dwarf mass and core chemical composition but otherwise occurs at a specific evolutionary stage, resulting in it being a possible explanation for the close clustering of DA(H)e stars on the \textit{Gaia} HRD \citep{Schreiber2021b}.

Core crystallization combined with the white dwarf's rotation has been predicted to sustain a global magnetic dynamo \citep{Ginzburg2022}. The dependence of these factors results in a relationship between the fraction of the core which is crystallized, the spin period and magnetic field strength \citep{Schreiber2021a, Schreiber2021b, Ginzburg2022, Fuentes2023}. We used evolutionary model sequences to calculate the fraction of the DAe stars which is crystallized based on their photometric parameters (see Section~\ref{sec:crystallization}). The core oxygen abundance in white dwarfs is poorly understood, thus uncertainty is introduced in our evolutionary model sequences as this parameter influences the onset of crystallization. To acknowledge this uncertainty, we calculated two sets of models: one with a standard core composition of 60 per cent oxygen (left panel of Figure~\ref{fig:DAe_crystallization}); and one with a heavier core composition of 80 per cent oxygen (right panel of Figure~\ref{fig:DAe_crystallization}). It is apparent that the onset of crystallization occurs earlier when a higher core oxygen abundance is assumed. 

From our evolutionary model sequences, we cannot determine whether WD\,J0412$+$7549 has started to crystallize or not as it depends on its core composition (which is uncertain) and the precision of its \Teff\ and $\log g$ photometric parameters. If WD\,J0412$+$7549 has an 80 per cent core oxygen abundance, the crystallized fraction could be as high as 40 per cent at $3\sigma$. We can exclude that WD\,J1653$-$1001 has started to crystallize considering up to $3\sigma$ of its photometric parameters and both 60 per cent and 80 per cent core oxygen abundances. 

The position of DA(H)e stars on the \textit{Gaia} HRD is compared to the onset of crystallization in Figure~\ref{fig:HRD}. In this work, we have improved the modelling in different ways and therefore we can now be relatively confident that the atmospheric parameters are robust. First, the photometric parameters (\Teff/$\log g$) of DA(H)e white dwarfs with 3D convective or 1D radiative models (accounting for magnetic effects) were shown to be similar (see Section~\ref{sec:parameters}). 
It has also recently been shown that photometric atmospheric parameters for cool $\approx 8000$\,K magnetic white dwarfs using non-magnetic models are likely to be accurate \citep{McCleery2020,Hardy2023}. 

In addition, the crystallization sequences we used include up-to-date physics (see Section~\ref{sec:crystallization}). When an oxygen mass fraction of 80 per cent is assumed, two out of 28 DA(H)e stars have not started to crystallize within 3$\sigma$. When an oxygen mass fraction of 60 per cent is assumed, out of the 28 DA(H)e white dwarfs seven have not started to crystallize within 1$\sigma$ nor five within 3$\sigma$. Uncertainties still remain in crystallization models \citep[e.g.][]{Blouin2021_distil}, but at present the predicted onset of crystallization does not fully match with the emergence of DA(H)e stars, and even then it is expected that the emergence of surface magnetic fields from a crystallization dynamo will be delayed from the onset of crystallization \citep{Ginzburg2022}.

An anti-correlation between the magnetic field strength and the white dwarf spin period is expected for a crystallization-driven dynamo \citep{Ginzburg2022}. The tentative positive correlation found by \citet{Manser2023} in the nine DAHe stars with ZTF-determined spin periods and the three previously published DAHe stars is not in agreement with the relationship $B \propto P^{-1/2}$ found by \citet{Ginzburg2022} for a magnetic field generated by a crystallization-driven convective dynamo in single white dwarfs. This observation, coupled with the fact that some DAe stars have likely not started to crystallize, makes it challenging to unequivocally attribute crystallization as the universal origin mechanism for their magnetic fields. However, emergence of magnetic fields from another mechanism such as magnetic diffusion of pre-white dwarf fields over Gyr timescales, can still be an explanation for DA(H)e stars \citep{Cantiello2016,BagnuloLandstreet2021,Bagnulo2022}, even though it needs fine tuning for the emergence to almost coincide with the onset of crystallization.

From studies of volume limited samples, magnetism has been found to be rare and generally weak in young canonical mass ($\approx 0.6$\,\Msun) white dwarfs, yet the magnetic incidence appears to increase for cooling ages larger than $2 - 3$\,Gyr \citep{Bagnulo2022}. This is likely evidence that magnetic fields emerge from the interior to the stellar surface within this age range. The delayed magnetic field emergence could explain the existence of DA(H)e stars and their close clustering on the \textit{Gaia} HRD, rather than scenarios where the magnetic field has always been at the white dwarf surface. Field emergence is not unique to any magnetic generation scenario so we cannot narrow down the origin of the fields. It is yet to be determined whether DA(H)e stars were originally DA where a mechanism caused the emergence of both the magnetic field and spectral emission features together, or if they were DAH stars which were incident to a mechanism which produces Balmer line emission. However, the discovery of delayed magnetic field emergence -- independent of DA(H)e observations -- gives support to the scenario that DA are the progenitors of DA(H)e stars, suggesting Balmer line emission is connected to the magnetic field generation mechanism(s). 

It is interesting to note that the DA(H)e variability strip in the \textit{Gaia} HRD coincides with the maximum strength of the hydrogen opacity in the non-degenerate envelope \citep[see fig.~\,6 of][]{Saumon2022}, and this opacity bump is even more sharply peaked in stellar radius for radiative magnetic structures \citep{Tremblay2015}. We speculate that it could result in instabilities and waves responsible for activity at the surface, but it would not explain a change in the incidence of magnetic white dwarfs with temperature \citep{Bagnulo2022}.

\subsection{An explanation for photometric variability in \texorpdfstring{WD\,J0412$+$7549}{WDJ0412+7549}}

In order to explain the causes of photometric variability, we used the \textit{TESS} bandpass and integrated different white dwarf model fluxes to perform two tests. In the first test, we computed the flux using our best-fitting 3D convective DA model atmosphere without line emission, and then with added artificial line emission based on the observed Keck spectrum at peak emission. The flux variation was 0.04 per cent, which rules out that Balmer line emission is causing the observed photometric variation of $\approx 2.29 \pm 0.05$ per cent. 

\begin{figure}
\centering
\includegraphics[width=\columnwidth]{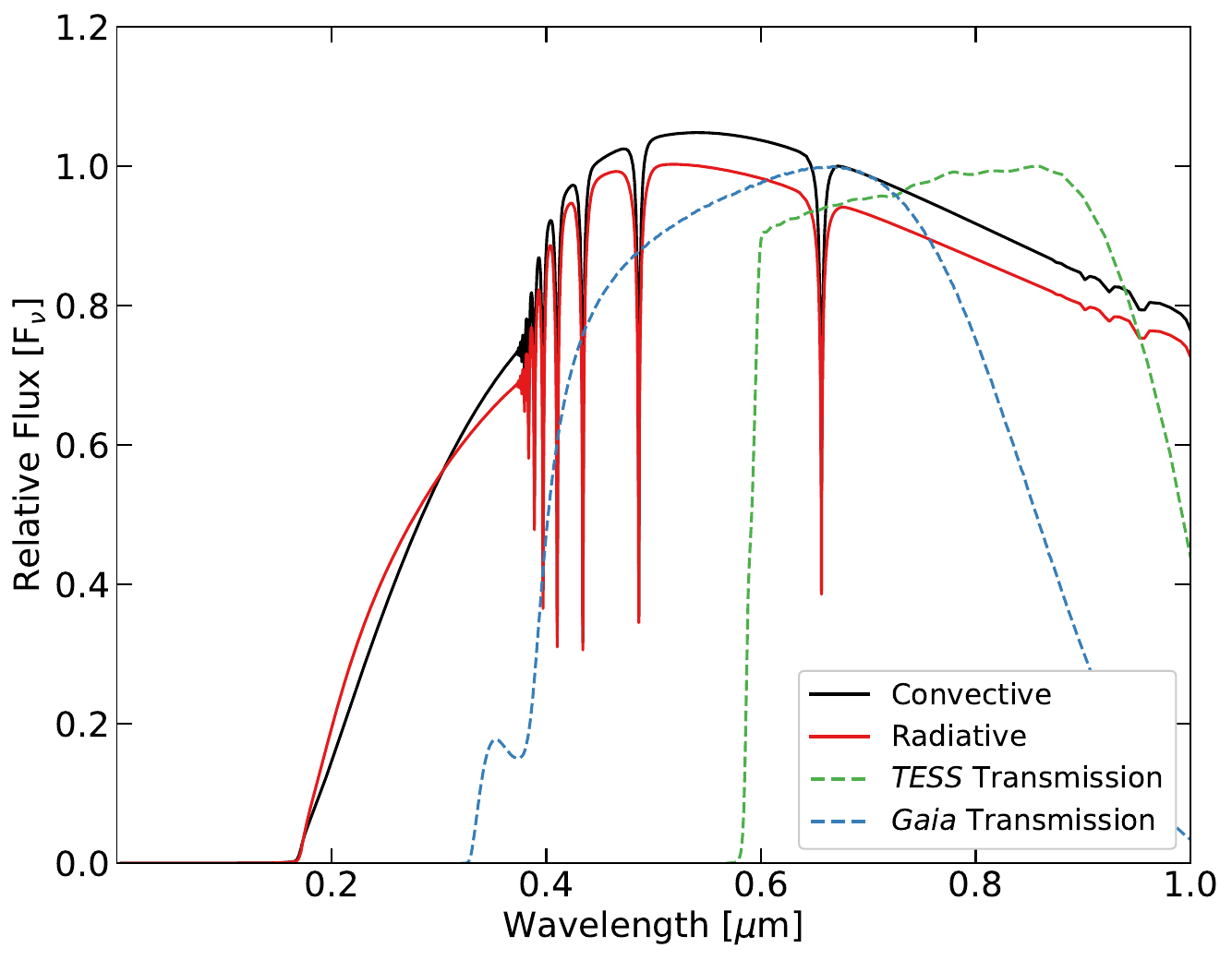}
    \caption{Model spectra created from the same 1D model atmosphere code with a pure-hydrogen equation-of-state, showing radiative (red curve) and convective (ML2/$\alpha$ = 0.8; black curve) structures corresponding to opposing sides of the star. A constant effective temperature across the white dwarf surface is assumed. The normalized \textit{TESS} (green) and \textit{Gaia} (blue) bandpasses are shown.}
\label{fig:photVar_test}
\end{figure}

In the second test, we used our 1D radiative and 1D convective DA model atmospheres\footnote{The use of 1D convective models is to allow for a better differential comparison using the same atmosphere code.} on opposing sides of the star to model convective and radiative magnetic regions, respectively (Figure~\ref{fig:photVar_test}). We assumed a constant effective temperature across the surface. The impact of magnetic fields on a stellar structure can be estimated from the plasma-$\beta$ parameter ($\beta = 8\pi P / B^2$, where $P$ is the thermal pressure and $B$ the magnetic field strength). It was found from 3D magneto-hydrodynamic (MHD) simulations \citep{Tremblay2015} that a value of $\beta \sim 1$ will inhibit convective energy transfer in the white dwarf atmosphere. Magnetic fields larger than this critical value, corresponding to $0.002 - 0.1$\,MG in DA white dwarfs \citep{Cunningham2020}, will therefore result in a largely radiative temperature stratification \citep{GF2018}. For WD\,J0412$+$7549, the critical field to inhibit convection is $B \approx 0.01$\,MG, so a plausible scenario to explain photometric variability is that one side of the star harbours a small magnetic field (above $B \approx 0.01$\,MG but below the detection limit of 0.05\,MG) resulting in a radiative structure, while other regions are able to remain convective \citep[][see also Section \ref{sec:parameters}]{Tremblay2015}.

The different atmospheric temperature stratifications in convective and magnetic regions result in different SEDs. We find that flux peaks on the convective side of the star in the \textit{TESS} bandpass, with a flux deficit on the radiative side. A flux variation of $\approx$ 5 per cent is predicted between the two regions, which suggests that even subtle effects on the atmospheric structure from small magnetic fields ($\approx 10 - 50$\,kG) could be responsible for the photometric \textit{TESS} flux variation. Under this scenario, the photometric minimum corresponds to the radiative (magnetic) side, in line with the expectation of an active chromosphere and line emission in magnetic regions. While the effective temperature remains constant, the smaller flux in the \textit{TESS} bandpass for the radiative region implies that light is emitted from a slightly cooler region of the atmosphere. This is analogous to the previously suggested scenario where DA(H)e stars have cool magnetic regions \citep{Walters2021}, but does not require a yet unexplained mechanism to change the effective temperature across the surface. 

We note that the above prediction could be tested with multi-wavelength photometric studies and spectropolarimetry, to confirm that one or more phases have a radiative structure and magnetic field strength $\approx 10 - 50$\,kG. We also note that this scenario is unlikely to apply for DAHe white dwarfs, in particular for GD\,356 where spectropolarimetric observations suggest that in all rotation phases the surface is highly magnetic \citep{Walters2021}, hence likely radiative.

\section{Conclusions}
\label{sec:Conclusions}

Two DAe white dwarfs, WD\,J0412$+$7549 and WD\,J1653$-$1001, have been discovered in the past three years. They have hydrogen-dominated atmospheres and exhibit H$\alpha$ and H$\beta$ line emission. These characteristics are the same as DAHe stars, yet DAHe stars have Zeeman-split Balmer emission lines indicating magnetism. The class DA(H)e incorporates DAe and DAHe stars due to the hypothesis that they have a similar origin, as their uncoincidental close clustering on the \textit{Gaia} HRD indicates these stars may be experiencing a short evolutionary phase where magnetic fields with vastly different strengths trigger Balmer line emission. The physical mechanism(s) causing DA(H)e stars is still unknown.

This work provides detailed follow-up of the two known DAe stars by presenting new time-domain spectroscopic observations and analysis of the latest photometric time-series data. Both DAe stars do not have detectable magnetic fields, with upper limits on magnetic field strength of $B < 0.05$\,MG. We confirm that both DAe stars exhibit photometric and spectroscopic variability, which we interpret as the white dwarf spin period. WD\,J0412$+$7549 has a period of $2.2891144 \pm 0.0000016$\,h and WD\,J1653$-$1001 has a tentative period of $80.534 \pm 0.087$\,h. Additional photometric follow-up with larger phase coverage is needed to confidently determine the period of WD\,J1653$-$1001 as the current ZTF data is contaminated by an unresolved background main-sequence star. 

WD\,J0412$+$7549 has spectroscopic data spanning its entire phase, including two high-resolution Keck spectra. We calculated the radial velocity of this star using its H$\alpha$ emission line and found an absence of significant radial velocity variation, which indicates it is originating from the white dwarf surface or chromosphere. The similar physical characteristics of DAe and DAHe stars (Section~\ref{sec:physical characteristics}) suggests that the mechanism causing the Balmer line emission and photometric variability is in common, and likely intrinsic to the white dwarf, rather than a substellar companion body closely orbiting the stars \citep{Walters2021}. Furthermore, WD\,J0412$+$7549 has an anti-phase relationship between photometric (flux) and spectroscopic (emission) variability which is the same phase relationship found in DAHe stars. An anti-phase relationship is the expectation of a photospheric dark spot/region with a temperature-inverted and optically thin chromospheric emission region, and not of a closely orbiting companion. 

We have shown that the photometric flux variation in WD\,J0412$+$7549 could be readily explained by radiative (magnetic) and convective hemispheres having different SEDs, with a threshold of $B \approx 10$\,kG between the two faces. Testing this scenario would require better phase resolved limits on magnetic fields down to $1-10$\,kG, a measurement that could be made difficult by the short rotation period.

It is possible that magnetic emergence has not occurred yet or has just started to occur in DAe stars, which explains their lack of a detectable magnetic field but similarities with DAHe stars. Therefore, the physical origin of their characteristics could be the same as DAHe stars. We explored magnetic mechanisms that could drive a magnetic field in isolated white dwarfs, including global and local surface dynamos. A global dynamo is created from the combination of core crystallization and the white dwarf rotation. Our modelling could not determine if WD\,J0412$+$7549 has started to crystallize but we conclude that WD\,J1653$-$1001 has not started to crystallize. Spectroscopic and multi-wavelength photometric follow-up of DAe stars is required to further understand their behaviours and to determine the origin of this class.

\section*{Acknowledgements}

This project has received funding from the European Research Council under the European Union’s Horizon 2020 research and innovation programme (Grant agreement numbers 101002408~--~MOS100PC and 101020057~--~WDPLANETS) as well as the UK STFC consolidated grant ST/T000406/1. The authors acknowledge financial support from Imperial College London through an Imperial College Research Fellowship grant awarded to CJM.

This work has made use of data from the European Space Agency (ESA) mission
{\it Gaia} (\url{https://www.cosmos.esa.int/gaia}), processed by the {\it Gaia} Data Processing and Analysis Consortium (DPAC, \url{https://www.cosmos.esa.int/web/gaia/dpac/consortium}). Funding for the DPAC has been provided by national institutions, in particular the institutions participating in the {\it Gaia} Multilateral Agreement.

This paper includes data collected by the \textit{TESS} mission. Funding for the \textit{TESS} mission is provided by the NASA's Science Mission Directorate. This paper includes data collected by the \textit{TESS} mission that are publicly available from the Mikulski Archive for Space Telescopes (MAST).

Based on observations obtained with the Samuel Oschin Telescope 48-inch and the 60-inch Telescope at the Palomar Observatory as part of the Zwicky Transient Facility project. ZTF is supported by the National Science Foundation under Grants No. AST-1440341 and AST-2034437 and a collaboration including current partners Caltech, IPAC, the Weizmann Institute for Science, the Oskar Klein Center at Stockholm University, the University of Maryland, Deutsches Elektronen-Synchrotron and Humboldt University, the TANGO Consortium of Taiwan, the University of Wisconsin at Milwaukee, Trinity College Dublin, Lawrence Livermore National Laboratories, IN2P3, University of Warwick, Ruhr University Bochum, Northwestern University and former partners the University of Washington, Los Alamos National Laboratories, and Lawrence Berkeley National Laboratories. Operations are conducted by COO, IPAC, and UW.

Some of the data presented in this work were obtained at the W. M. Keck Observatory, which is operated as a scientific partnership among the California Institute of Technology, the University of California and the National Aeronautics and Space Administration. The Observatory was made possible by the generous financial support of the W. M. Keck Foundation. We wish to recognize and acknowledge the very significant cultural role and reverence that the summit of Maunakea has always had within the indigenous Hawaiian community. We are most fortunate to have the opportunity to conduct observations from this mountain.

The William Herschel Telescope and Isaac Newton Telescope are operated on the island of La Palma by the Isaac Newton Group of Telescopes in the Spanish Observatorio del Roque de Los Muchachos of the Instituto de Astrofísica de Canarias.

Based on observations obtained at the international Gemini Observatory, a programme of NSF’s NOIRLab, which is managed by the Association of Universities for Research in Astronomy (AURA) under a cooperative agreement with the National Science Foundation on behalf of the Gemini Observatory partnership: the National Science Foundation (United States), National Research Council (Canada), Agencia Nacional de Investigaci\'{o}n y Desarrollo (Chile), Ministerio de Ciencia, Tecnolog\'{i}a e Innovaci\'{o}n (Argentina), Minist\'{e}rio da Ci\^{e}ncia, Tecnologia, Inova\c{c}\~{o}es e Comunica\c{c}\~{o}es (Brazil), and Korea Astronomy and Space Science Institute (Republic of Korea).

Research at Lick Observatory is partially supported by a generous gift from Google. A major upgrade of the Kast spectrograph on the Shane 3 m telescope at Lick Observatory was made possible through generous gifts from William and Marina Kast as well as the Heising-Simons Foundation.

\section*{Data Availability}
Spectroscopy from the WHT, Keck, INT, Gemini and Shane telescopes, in addition to photometry from ZTF, are available from their respective public archives. The \textit{TESS} data are accessible via the MAST (Mikulski Archive for Space Telescopes) portal at \href{https://mast.stsci.edu/portal/Mashup/Clients/Mast/Portal.html}{https://mast.stsci.edu/portal/Mashup/Clients/Mast/Portal.html}.



\bibliographystyle{mnras}
\bibliography{all} 




\appendix

\section{Time-domain spectra of \texorpdfstring{WD\,J0412$+$7549}{WDJ0412+7549}}

\begin{figure*}
	\includegraphics[width=2\columnwidth]{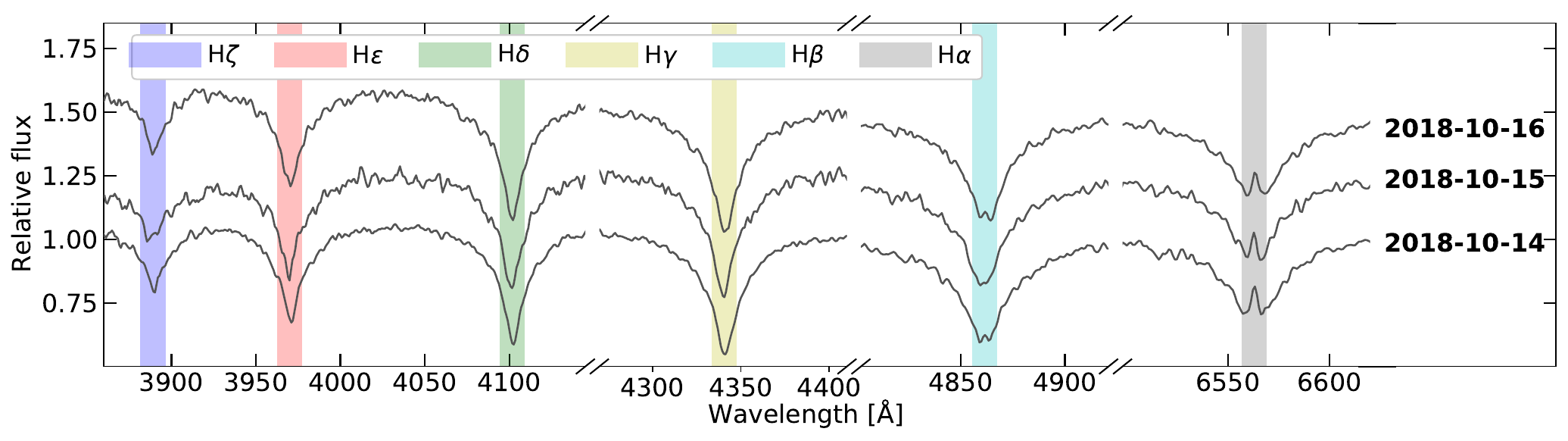}
    \caption{WHT spectra from three consecutive nights of observations of WD\,J0412$+$7549, taken around the H$\alpha$ to H$\zeta$ line regions. Three exposures were taken on 2018 October 14 however we show the stacked spectrum here. The observation UT dates are shown on the right of the plot. Spectra are convolved with a Gaussian with a FWHM of 2\,\AA\ and offset vertically for clarity.}
    \label{fig:WHT_Ha_Hb_Hg_Hd_Hz}
\end{figure*}

\begin{figure*}
	\includegraphics[width=1.8\columnwidth]{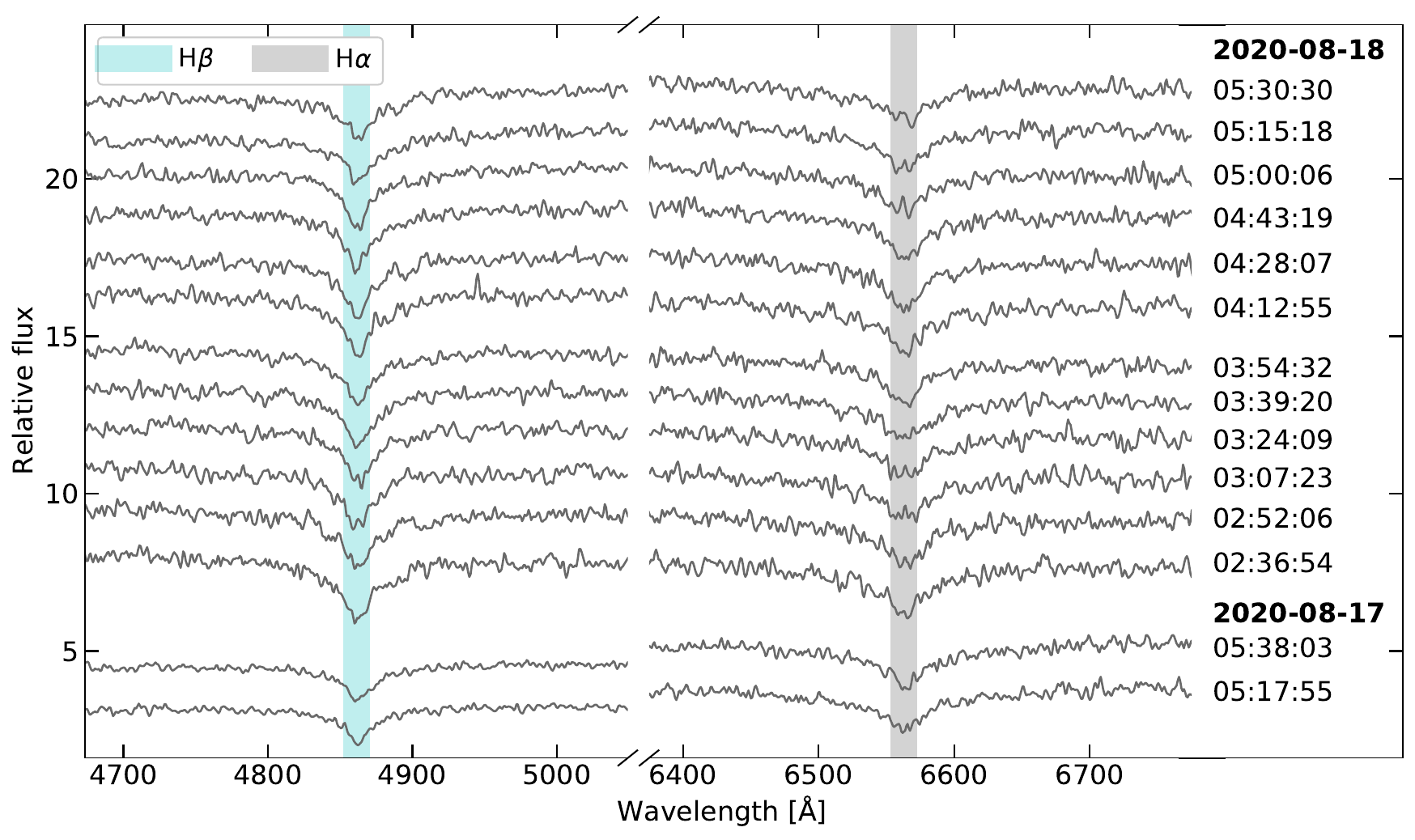}
    \caption{14 exposures of INT IDS spectroscopy taken around the emission core of H$\alpha$ and H$\beta$. The observation UT date and start times are shown on the right of the plot. Spectra are convolved with a Gaussian with a FWHM of 2\,\AA\ and offset vertically for clarity.}
    \label{fig:INT_Ha_Hb}
\end{figure*}

\begin{figure*}
	\includegraphics[width=2\columnwidth]{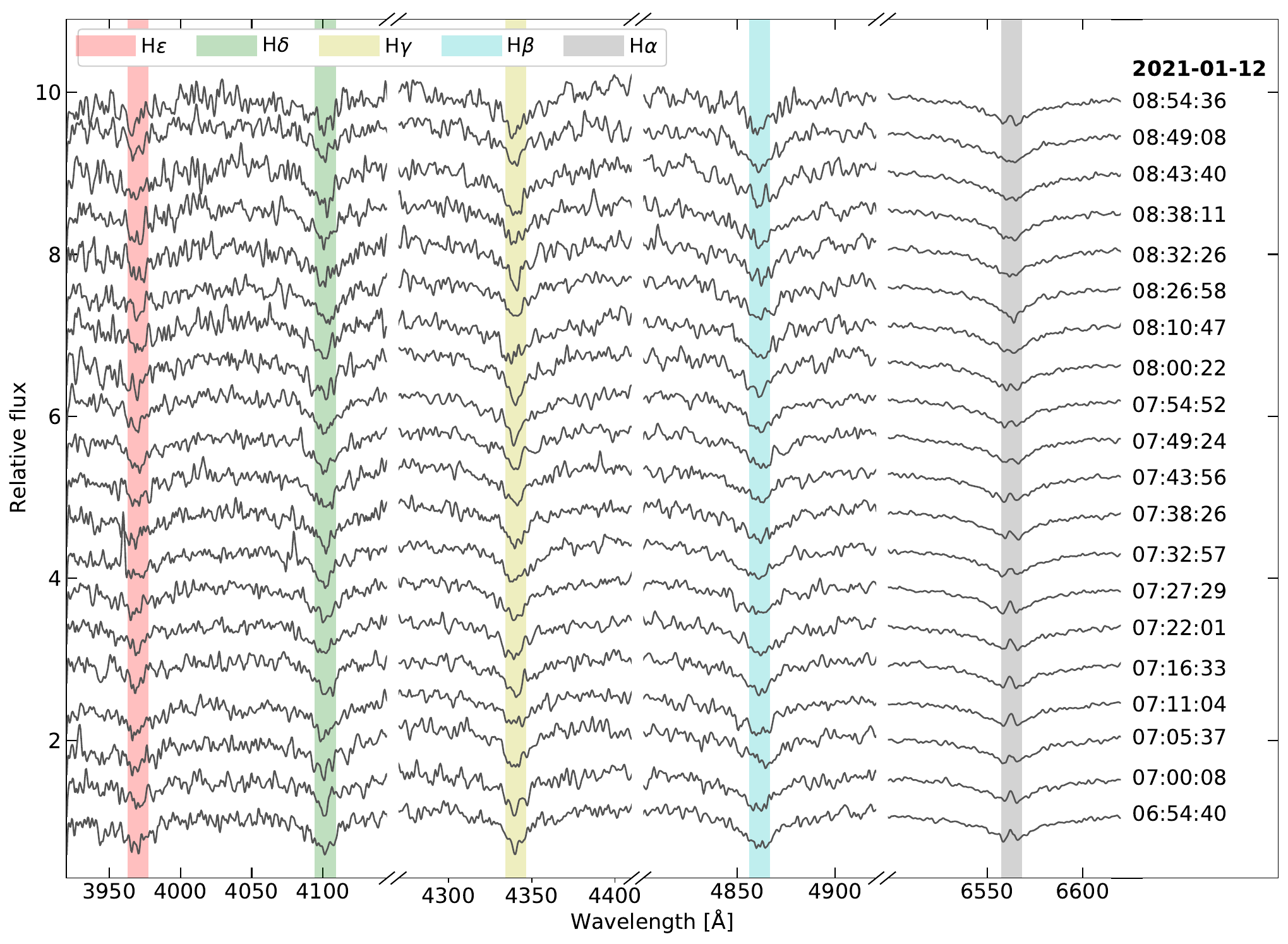}
    \caption{Gemini spectra of the 20 exposures taken on 2021 January 12 of WD\,J0412$+$7549 around the H$\alpha$ to H$\delta$ line regions. The observation UT date and start times are shown on the right of the plot. Spectra are convolved with a Gaussian with a FWHM of 3\,\AA\ and offset vertically for clarity.}
    \label{fig:gemini_Ha_Hb_Hg_Hd}
\end{figure*}


\bsp	
\label{lastpage}
\end{document}